\documentclass[sageV,times]{sagej}
\makeatletter \oddsidemargin 0.0cm \evensidemargin 0.0cm
\usepackage{moreverb,url}
\usepackage{float}
\usepackage{graphicx,psfrag,subfigure,epstopdf}
\usepackage[normalem]{ulem}
\usepackage{moreverb,url}
\usepackage{amssymb}
\usepackage[square,sort&compress,numbers]{natbib}
\usepackage[colorlinks,bookmarksopen,bookmarksnumbered,citecolor=red,urlcolor=red]{hyperref}

\newcommand\BibTeX{{\rmfamily B\kern-.05em \textsc{i\kern-.025em b}\kern-.08em
T\kern-.1667em\lower.7ex\hbox{E}\kern-.125emX}}

\usepackage{amsmath,amsfonts,amssymb}
\usepackage{stmaryrd}
\usepackage{grffile}
\usepackage{natbib}
\usepackage{graphicx}
\usepackage{subfigure}
\usepackage{color}
\usepackage{textpos}
\usepackage{mathrsfs}
\usepackage{smartref}
\usepackage{bm}

\def \Div{\mbox{Div\hskip 1pt}}
\def \tr{\mbox{tr\hskip 1pt}}

\begin{document}

\runninghead{One-dimensional model}

\title{Reduced model and nonlinear analysis of localized instabilities of residually stressed cylinders under axial stretch}

\author{Yang Liu\affilnum{1,2}, Xiang Yu\affilnum{3},  Luis Dorfmann\affilnum{4}}

\affiliation{\affilnum{1}Department of Mechanics, School of Mechanical Engineering, Tianjin University, Tianjin 300354, China. \\
\affilnum{2}Tianjin Key Laboratory of Modern Engineering Mechanics, Tianjin 300354, China.\\
\affilnum{3}{Department of Mathematics, School of Computer Science and Technology, Dongguan University of Technology, Dongguan, 519087, China}.\\
\affilnum{4}Department of Civil and Environmental Engineering, Tufts University, Medford, MA 02155, USA.}

\corrauth{Luis Dorfmann} 
\email{Luis.Dorfmann@tufts.edu}

\begin{abstract}
In this paper we present a dimensional reduction to obtain a one-dimensional model to analyze localized necking or bulging in a residually stressed circular cylindrical solid. The  nonlinear theory of elasticity is first specialized to obtain the equations governing the homogeneous deformation.  Then, to analyze the non-homogeneous part, we include higher order correction terms of the axisymmetric displacement components leading to a three-dimensional form of the total potential energy functional. Details of the  reduction to the one-dimensional form are given. We focus on a residually stressed Gent material and use numerical methods to solve the governing equations. Two loading conditions are considered. In the first, the residual stress  is maintained constant, while the axial stretch is used as the loading parameter. In the second, we keep the pre-stretch  constant and monotonically increase the residual stress until bifurcation occurs. We specify initial conditions, find the critical values for localized bifurcation and compute the change in radius during  localized necking or bulging growth. Finally, we optimize material properties and use the one-dimensional model to simulate necking or bulging until the Maxwell values of stretch are reached. 

\end{abstract}


\maketitle

\section{Introduction}\label{Introduction}

In 1890, Mallock \cite{Mallock1890}   described the sequence of events leading up to localized bulging of thin-walled India-rubber (natural rubber) tubes subject  to internal fluid pressure. In particular, he observed that the tube maintains its cylindrical form until the increase in radius remains proportional to the reference radius. But, when more fluid is introduced the tube becomes unstable and the internal pressure diminishes. 

When more fluid is introduced then necessary to expand to its stable limit, the tube no longer remains cylindrical throughout its length, but assumes the form of a cylinder with one or more bulges. He observed that the diameter of the part which remains cylindrical, though greater than the reference diameter, is less than that attained at the stable limit. This is the first documented experiment in which bulging  is induced by internal pressure in thin-walled rubber tubes, but it is only in recent years that it has been recognized that such nonlinear phenomena could be used in industrial applications. 

Ogden \cite{Ogden1997} develops the theory of  incremental deformation superimposed on a known finitely deformed configuration. Linear approximations in terms of the incremental deformation and its gradient are used to examine incremental constitutive laws and equilibrium equations. In the theory presented in this paper, the incremental displacement components contain higher order correction terms,  allowing for a nonlinear stability analysis and for the evolution of localized necking or bulging in prismatic solids. A summary of the developments of nonlinear stability analysis of thick  hyperelastic plates is given  in \cite{Fu1999}.  First, the linear theory is used for two representative problems to determine the critical values for stability. The linear theory can determine the buckling mode, but not its amplitude, which can only be obtained from a fully nonlinear analysis. On the other hand, the weakly nonlinear analysis is concerned with the mode amplitude when the applied load deviates from its critical value by a small amount with  the amplitude depending on a far distance variable. 

Methods of nonlinear stability analysis of elastic bodies are summarized in  \cite{Fu2001}. Particular attention is  on using the perturbation approach to the stability analysis of elastic bodies subject to large deformation. Two types of buckling modes are considered, the first consists of stability studies in which buckling  modes are periodic, the second  focuses on bifurcations  where the critical mode number is set to zero.

The membrane assumption is used in \cite{Fu2008} to analyze the bifurcation conditions of an infinitely long hyperelastic tube with the axial stretch maintained uniform at infinity, while inflated by internal pressure. The bifurcation conditions and the near-critical behavior are determined analytically. It is shown that the bifurcation mode having zero wave number correctly describes localized bulging and necking initiation, which grow and evolve into a two-phase deformation. The near-critical post-bifurcation analysis is used to derive the amplitude equation, which is used to determine under what conditions a localized solution is possible. A more realistic case of an inflated  tube with closed ends is considered in \cite{Pearce2010}. There, the entire bulging or necking process is determined, from initiation to the fully nonlinear propagation and the stability properties of the weakly nonlinear  and fully nonlinear  solutions are evaluated.  

The use of the membrane theory to assess localized bulging in cylindrical tubes with finite wall thickness may not lead to accurate results. This  is addressed in  \cite{Fu2016}, where the  theory of nonlinear elasticity is used to derive an explicit bifurcation condition for localized bulging in a  hyperelastic tube of arbitrary thickness subjected to internal pressure and axial extension. The condition  requires that the Jacobian determinant of the internal pressure and resultant axial force, each considered a function of the two principal stretches, vanishes.  Further insight into the bifurcation condition derived in \cite{Fu2016} is given in \cite{Yu2022}, where it is applied to derive the localized condition of an inflated hyperelastic tube and to the axisymmetric  necking of a thin sheet subject to equibiaxial stretching \cite{Fu2018}. 

For thin-walled tubes the conservation laws  reduce to ordinary differential equation and when the membrane assumption is no longer applicable,  these become  nonlinear partial differential equations.  Even the weakly nonlinear near-critical analysis  is  no longer trivial, but can be performed using asymptotic methods, while the fully nonlinear post-buckling behavior relies on finite element techniques \cite{Ye2020}. 

Experimental data  are used in  \cite{Wang2019} to validate the  theoretical prediction of  the initiation and propagation  pressures  in  rubber tubes with different lengths,  wall thicknesses and  end conditions. It is found that  for constant axial force the initiation pressure is given by a bifurcation condition and the propagation pressure is determined by  Maxwell's equal-area rule. The initiation pressure for the fixed-ends case, where the length of the tube is fixed once the pre-stretch has been applied, is again determined by a bifurcation condition, but the propagation pressure is no longer determined by Maxwell's equal-area rule. The pressure, following bulge initiation decreases to a minimum and then rises again, the latter accompanied by bulge propagation. The experimental results of the inflated tube with fixed ends are also used in \cite{Guo2022} to validate an analytical model.

In  \cite{Audoly2016} a dimensional reduction is proposed to obtain a one-dimensional energy function, which is used to analyze the fully nonlinear evolution of  tensile necking in  prismatic solids having an arbitrary cross-section. This is made possible by incorporating the dependence on the axial stretch and its gradient that arise during necking.  In  \cite{Lestringant2018} the  method is specialized to obtain a one-dimensional diffuse interface model, which is then used to analyze localized bifurcation in nonlinear axisymmetric membrane tubes. A systematic expansion of the membrane energy in terms of the aspect ration $\varepsilon=R/L\leqslant 1$ is obtained, where $R$ and $L$ are the initial radius and length, respectively. In \cite{Audoly2019} the formulation is generalized and then applied to analyze necking of rate dependent  axially stretched round bars and to necking in thin sheet materials.  A systematic dimensional reduction  is proposed in \cite{Lestringant2020JMPS} resulting in a nonlinear structural model of prismatic  solids, which again captures the strain gradient effect. The one-dimensional model allows for large and inhomogeneous changes in the shape of the cross-section and can be used to analyze localization in slender structures. In \cite{Lestringant2020PRSA} the one-dimensional strain gradient model is used to predict necking in cylindrical nonlinear elastic  solids generated by  an increase of the surface tension.  The dimensional reduction methodology is adopted in  \cite{Yu2023} to obtain a one-dimensional model for the analysis of bulging or necking in an inflated hyperelastic tube with arbitrary wall thickness. The theory is  applied in two limiting cases, one investigates bulging in a membrane tube, the other necking in a stretched solid cylinder. It is shown that the  one-dimensional model is capable to describe the entire bulging evolution accurately as demonstrated by  comparing analytical results with finite element simulations.

In this paper we derive a one-dimensional model to analyze localized necking or bulging in a residually stressed circular cylindrical solid. This study builds upon our results in \cite{Liu2023MMS}, where we investigated localized necking or bulging in a residually stressed  solid cylinder with circular cross-sectional area  subject to axial stretch. We used the theory of linearized incremental deformations and presented the governing equations and boundary conditions in Stroh form. The results show that  following bifurcation,  bulged and necked regions  connected by transition zones propagate along the axial direction of the cylinder.  

The paper is organized as follows. In Section \ref{Three-dimensional} we formulate the three-dimensional problem and provide an explicit expression of the axial force to maintain the cylindrical form of a residually stressed  solid. In Section \ref{1D-model} we use a dimensional reduction to obtain a one-dimensional model for the nonlinear analysis of necking and bulging. A key assumption of the theory is that  the bifurcation mode varies slowly in the axial direction. The axisymmetric incremental deformation components are expanded to include higher-order correction terms. A dimensional reduction is then used to derive a one-dimensional energy functional. Minimization of the energy with respect to the stretch $\lambda$ results in a second order nonlinear differential equation. To obtain numerical results we define  in Section \ref{Material model} an  incompressible Gent material and  the explicit forms of the residual stress components. The dependence  of the critical stretch and residual stress values  on the material extensibility is shown.  The focus in Section  \ref{Nonlinear} is on the use of the one-dimensional model to simulate initiation and growth of a local bifurcation. Two loading sequences are considered. In Section \ref{Fixed-stress} the residual stress is maintained constant and the axial stretch is used as the loading parameter. In Section \ref{Fixed-length} the constant pre-stretch is applied first and the residual stress is then increased monotonically until bifurcation occurs. Concluding remarks are given in Section \ref{Conclusions}.

\section{Three-dimensional formulation and primary  deformation}\label{Three-dimensional}

A linear bifurcation analysis was used  in \cite{Liu2023MMS}  to investigate localized necking and bulging in a residually stressed  solid cylinder with circular cross-sectional area  subject to axial stretch.  The method was based on the theory of linearized incremental deformations superimposed on a known finitely deformed configuration with the governing equations and boundary conditions   given in Stroh form. Following bifurcation,   bulged and necked regions develop connected by transition zones, which  simply propagate along the axial direction of the cylinder.  In the current paper we use a dimensional reduction to obtain a one-dimensional model for the analysis of bulging or necking in a stretched  residually stressed cylindrical solid. The derived model is used to investigate the post-bifurcation behavior in the fully nonlinear regime. 

\begin{figure}[!ht]
	\centering\includegraphics[scale=0.8]{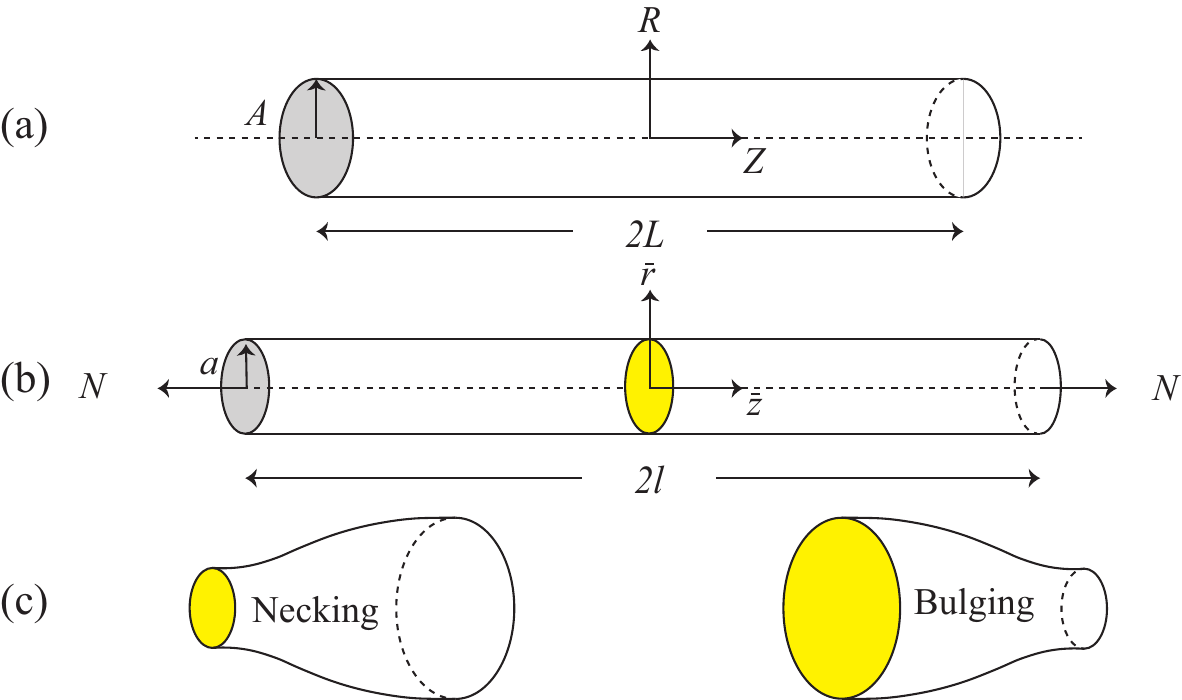}\caption{(a) A hyperelastic cylindrical solid  with referential radius $A$, length $2L$ and zero traction on the curved surface; (b) The axial force $N$ first generates a homogeneous deformation with uniform stretch; (c) With increasing deformation, depending on the specific loading scenario,  localized necking or bulging may occur, see  \citep{Liu2023MMS}.}\label{geometry}
\end{figure}

We  consider the application of the one-dimensional theory to investigate  bifurcation  of  a  circular cylindrical solid with reference geometry defined in cylindrical polar coordinates $(\Theta, Z, R)$ by
\begin{equation}
\label{RefConfig}
 0\leqslant\Theta \leqslant 2 \pi,\quad  -L \leqslant Z \leqslant L,\quad 0\leqslant R\leqslant A,
\end{equation}
where  $A$ is the outer radius and $2 L$ the initial total length, see Figure \ref{geometry}(a). The axial extension of the cylinder is achieved by the application of an axial force $N$ resulting in a homogeneous   or inhomogeneous axisymmetric deformation. If the deformation is homogeneous,  the current geometry is given by 
\begin{equation}
\label{CurConfig}
0\leqslant\theta \leqslant 2 \pi,\quad  -l \leqslant z \leqslant l,\quad 0\leqslant r\leqslant a,
\end{equation}
where $(\theta, z,r)$ are the deformed cylindrical polar coordinates, and $a$ and $2 l$ are the outer radius and total deformed length, respectively. We use the standard basis vectors $(\mathbf e_{\theta},\mathbf e_z,\mathbf e_r)$ associated with both  $(\Theta, Z, R)$ and $(\theta, z,r)$. 

We focus attention to an incompressible material and describe a general axisymmetric deformation in the form
\begin{equation}
\label{Deformation}
\theta=\Theta,\quad z=z(Z,R),\quad r=r(Z,R),
\end{equation}
which encompasses the  homogeneous and inhomogeneous cases. The corresponding deformation gradient  $\mathbf{F}$ has the form 
\begin{equation}
\label{Deformation-gradient}
\mathbf{F}=\frac{r}{R}\mathbf e_\theta\otimes \mathbf e_\theta+z_Z\mathbf e_z\otimes \mathbf e_z+z_R\mathbf e_z\otimes \mathbf e_r+r_Z\mathbf e_r\otimes \mathbf e_z+r_R \mathbf e_r\otimes \mathbf e_r,
\end{equation}
where the subscripts $Z,R$ without a preceding comma indicate the corresponding partial derivatives. The incompressibility constraint requires that
\begin{equation}
\label{Incompressibility}
\det\mathbf{F}=1.
\end{equation}
It then follows from \eqref{Deformation-gradient} that 
\begin{equation}
    z_Zr_R-z_Rr_Z=0.
\end{equation}

There are no  mechanical  forces applied to the curved surface. Therefore, in the reference configuration  the residual stress  $\boldsymbol \tau$ satisfies the traction-free boundary condition
\begin{equation}
\label{Bc-residual-stress}
\boldsymbol \tau \mathbf N =\boldsymbol 0 \quad \mathrm{on} \quad R=A,
\end{equation}
where $\mathbf{N}$ denotes the  outward pointing unit vector normal to the surface $R=A$.  With no intrinsic couple stresses, the components of the stress $\boldsymbol \tau$ must satisfy  the equilibrium equation 
\begin{equation}
\label{Eq-residual-stress}
\Div \boldsymbol \tau=\boldsymbol 0,
\end{equation} 
where $\Div$ is the divergence operator defined in the reference configuration.

Consider a hyperelastic and incompressible material, where the mechanical properties are specified  in terms of an energy  function $W$ defined per unit reference volume. We follow the theory in \cite{Liu2023MMS} and consider an energy density, which is a function of the deformation gradient  $\mathbf F$ and of the residual stress $\boldsymbol \tau$ and takes the form $W(\mathbf F, \boldsymbol \tau)$. Rajagopal and Wineman \cite{Rajagopal2024}  show that a body with initial stresses/pre-stresses/residual stresses cannot be isotropic in such a configuration in which it is pre-stressed.  Reference is made to  \cite{Hoger1993},  where a general constitutive theory is developed, which is appropriate for hyperelastic, residually stressed, transversely isotropic material subject to large deformation.

The effect of the residual stress on the constitutive function is analogous to that of a structure tensor associated with preferred directions in fiber-reinforced materials, see \cite{Holzapfel2010}, for example.  Therefore, the residually stressed material is inhomogeneous and the mechanical properties anisotropic. For the considered  circular cylindrical  solid we assume that  the only nonzero components of $\boldsymbol \tau$ are $\tau_{\Theta\Theta}$ and $\tau_{RR}$  such that 
\begin{equation}
\label{Tau-exp}
\boldsymbol \tau=\tau_{\Theta\Theta}\mathbf e_\theta\otimes\mathbf e_\theta+\tau_{RR}\mathbf e_r\otimes\mathbf e_r,
\end{equation} 
which must satisfy the zero traction boundary condition \eqref{Bc-residual-stress} and  the equilibrium equation \eqref{Eq-residual-stress}.  The preferred direction generated by  \eqref{Tau-exp}  coincides with the axial direction of the cylinder and the material is therefore isotropic in the plane perpendicular to this direction. Transverse isotropy is the simplest material symmetry that a material can possess and still support a residual stress, see \cite{Hoger1985,Hoger1993} for details.

\subsection{Homogeneous solution }\label{Homogeneous-deformation}

In Figure \ref{geometry}(b) the magnitude of the applied  force $N$ is below a critical value resulting in a uniform axial stretch and a radial deformation. It follows that the deformation \eqref{Deformation} specializes to
\begin{equation}
\label{Deformation-homogeneous}
\bar{\theta}=\Theta,\quad \bar{z}=\lambda Z,\quad \bar{r}=\lambda^{-1/2}R,
\end{equation}
where $\lambda=l/L$ is  the constant axial stretch and the superposed bar  indicates   quantities associated with the homogeneous solution.  The  deformation gradient related to \eqref{Deformation-homogeneous} is then obtained as
\begin{equation}
\label{Deformation-gradient-homogeneous}
\bar{\mathbf{F}}=\lambda^{-1/2}\mathbf e_\theta\otimes \mathbf e_\theta+\lambda \mathbf e_z\otimes \mathbf e_z+\lambda^{-1/2}\mathbf e_r\otimes \mathbf e_r,
\end{equation}
resulting in  the three principal stretches
\begin{equation}
\label{Stretch-homogeneous}
\lambda_1=\lambda^{-1/2},\quad \lambda_2=\lambda,\quad \lambda_3=\lambda^{-1/2},
\end{equation}
where the subscripts 1, 2, 3 are used to indicate the $\theta$-, $z$-, $r$-directions, respectively. An increase in the axial force $N$ above a critical value will generate a bifurcation from the homogeneous cylindrical solution resulting in localized instabilities, see Figure \ref{geometry}(c).

We now restrict attention to a model where the residual stress  components  $\tau_{\Theta\Theta}$ and $\tau_{RR}$ are  functions of $R$ only. Because there are only three principal stretches in $\bar{\mathbf F}$ it is convenient to write the energy function $W$ in the reduced form as a function of $\lambda_1,\lambda_2,\lambda_3$ and $R$ given by
\begin{equation}
\label{Red-Energy}
W(\bar{\mathbf F},\boldsymbol\tau)=\phi(\lambda_1,\lambda_2,\lambda_3,R),
\end{equation}
with \eqref{Stretch-homogeneous} on the right hand side. Then, the nonzero components of the nominal stress tensor $\bar{\mathbf T}$  are  calculated by
\begin{equation}
\label{Stress-components}
\bar{T}_{ii}=\phi_i-\lambda_i^{-1}\bar{p},\quad \mathrm{no~summation},
\end{equation}
where $\phi_i = \partial \phi/\partial\lambda_i, i=1,2,3$ and $\bar{p}$ is a Lagrange multiplier enforcing the incompressibility condition \eqref{Incompressibility}. For an overview of the main constituents of the nonlinear theory of residually stressed materials the interested reader is referred to  \cite{Liu2023MMS,Melnikov2021,Melnikov2022}.

For the circular cylindrical geometry with uniform axial stretch the equilibrium equation $\Div \bar{\mathbf T}=\mathbf 0$ reduces to the single component in the radial direction 
\begin{equation}
\label{Eq-homogeneous}
\frac{\mathrm{d}\bar{T}_{33}}{\mathrm{d} R}+\frac{\bar{T}_{33}-\bar{T}_{11}}{R}=0,
\end{equation}
and the zero traction on the curved surface  $R=0$ gives rise to
\begin{equation}
\bar{T}_{33}=\int_A^{R}\frac{\phi_1-\phi_3}{t}\mathrm{d} t.
\end{equation} 
The boundary conditions are used to determine the value of the Lagrange multiplier $\bar{p}$ in \eqref{Stress-components}, but we omit the details for brevity.

The equilibrium condition in the axial direction requires that
\begin{equation}
N=2\pi\int_0^A\bar{T}_{22}R \mathrm d R,
\label{Force}
\end{equation}
and using \eqref{Stress-components}, \eqref{Eq-homogeneous} that \eqref{Force} this results in
\begin{equation}
N=\pi\int_0^A\left(2\phi_2-\lambda^{-3/2}\phi_1-\lambda^{-3/2}\phi_3\right)R\mathrm{d}R.
\end{equation}

\section{Derivation of a one-dimensional strain-gradient model}\label{1D-model}

In this section, we perform a proper dimensional reduction to derive a one-dimensional  model for the deformation and bifurcation of a residually stressed hyperelastic and incompressible cylinder.  A key ingredient of the dimensional reduction methodology in \cite{Audoly2016} is the fundamental assumption that the bifurcation mode varies slowly in the axial direction, which we assume to be the case for localized necking and bulging. Therefore, in deriving the one-dimensional model, it is convenient to introduce a variable $S$ defined by
\begin{equation}
\label{far-distance}
S=\varepsilon Z,
\end{equation}
where $\varepsilon$ is a small parameter, which in our case is taken as the radius to length ratio $A/L$. The reason we introduce the variable $S$ is to identify terms of different order $\varepsilon$ in the following derivation.

We begin the analysis by defining the total potential energy  $\mathcal E$ of a residually stressed, axisymmetrically deformed hyperelastic cylinder subject to a constant axial force $N$  by 
\begin{equation}
\label{Energy}
\mathcal{E}=2\pi\int_{-L}^{L}\int_0^{A} \left(W(\mathbf{F},\boldsymbol\tau)-\hat N z_Z\right)R\mathrm d R\mathrm d Z,
\end{equation}
where the force density $\hat{N}=N/(\pi A^2)$. In the following the  stretch $\lambda$ is no longer uniform and is taken as a function of the variable $S$. Following \cite{Yu2023}  we  augment the  axisymmetric displacement components \eqref{Deformation} to include the higher-order correction terms $\tilde w,\tilde u$ and look for an asymptotic solution of the form
\begin{eqnarray}
z(Z,R)&=&\dfrac{1}{\varepsilon}\int_0^S \lambda(\hat{S})\mathrm{d}\hat{S}+\varepsilon \tilde{w}(S,R)+\mathcal{O}(\varepsilon^3),\label{Expansion-z}\\
r(Z,R)&=&\lambda(S)^{-1/2}R+\varepsilon^2 \tilde{u}(S,R)+\mathcal{O}(\varepsilon^4),\label{Expansion-r}
\end{eqnarray}
where the stretch $\lambda(S)$ can be used for the homogeneous as well as for the bifurcated solution. The updated form of the deformation gradient \eqref{Deformation-gradient-homogeneous} is then given by
\begin{equation}
\label{Deformation-gradient-homogeneous-2}
\bar{\mathbf{F}}=\lambda(S)^{-1/2}\mathbf e_\theta\otimes \mathbf e_\theta+\lambda(S) \mathbf e_z\otimes \mathbf e_z+\lambda(S)^{-1/2}\mathbf e_r\otimes \mathbf e_r.
\end{equation}
Using the expansions  \eqref{Expansion-z} and \eqref{Expansion-r} in the general form of the deformation gradient \eqref{Deformation-gradient} gives
\begin{equation}
\label{Deformation-gradient-exp}
\mathbf{F}=\left[
\begin{array}{ccc}
\lambda(S)^{-1/2}+\dfrac{\tilde{u}}{R}\varepsilon^2& 0 & 0 \\
0 & \lambda(S)+\varepsilon^2\tilde{w}_{S} & \varepsilon \tilde{w}_{R} \\
0& -\dfrac{1}{2}\varepsilon\lambda^{-3/2}\lambda'(S)R & \lambda(S)^{-1/2}+\varepsilon^2\tilde{u}_{R} 
\end{array}
\right],
\end{equation}
where terms of $\mathcal{O}(\varepsilon^3)$ and higher have been neglected. The subscripts $S,R$ without a preceding comma indicate the corresponding partial derivatives, while differentiation of $\lambda(S)$ is  indicated by $\lambda'(S)=\mathrm{d}\lambda(S)/\mathrm{d}S$. Using \eqref{far-distance} results in the deformation gradient $\mathbf{F}$ being  a function of the  parameter $\varepsilon$. 

The next step in deriving the one-dimensional model is to expand the energy density function $W(\mathbf{F},\mathbf\tau)$  in terms of $\varepsilon$. This has the form
\begin{equation}
\label{Expansion-W}
W(\mathbf{F},\mathbf\tau)=W(\bar{\mathbf{F}},\mathbf\tau)+\mathbf{\Gamma}:(\mathbf{F}-\bar{\mathbf{F}})+\frac{1}{2}(\mathbf{F}-\bar{\mathbf{F}}):\boldsymbol {\mathcal A}:(\mathbf{F}-\bar{\mathbf{F}})+\mathcal{O}(\varepsilon^3), 
\end{equation}
where $\mathbf{\Gamma}$ and $\boldsymbol {\mathcal A}$ are defined by
\begin{equation}
\Gamma_{\alpha i}=\frac{\partial W}{\partial F_{i\alpha}}{\Bigg|_{\mathbf{F}=\bar{\mathbf{F}}}},\quad \mathcal{A}_{\alpha i \beta j}=\frac{\partial^2 W}{\partial F_{i\alpha}F_{j\beta}}{\Bigg|_{\mathbf{F}=\bar{\mathbf{F}}}},
\end{equation}
and for the double contraction we use the convention $\mathbf{\Gamma}:(\mathbf{F}-\bar{\mathbf{F}})=\Gamma_{\alpha i}(F_{i\alpha}-\bar{F}_{i\alpha})$. The component form of \eqref{Expansion-W} is then given by
\begin{equation}
W(\mathbf{F},\mathbf\tau)=W(\bar{\mathbf{F}},\mathbf\tau)+\Gamma_{\alpha i}(F_{i\alpha}-\bar{F}_{i\alpha})+\frac{1}{2}\mathcal{A}_{\alpha i \beta j}(F_{i\alpha}-\bar{F}_{i\alpha})(F_{j\beta}-\bar{F}_{j\beta})+\mathcal{O}(\varepsilon^3),
\label{Expansion-W-component}
\end{equation} 
with the major symmetries $\mathcal A_{\alpha i \beta j}=\mathcal A_{\beta j \alpha i}$ satisfied. It can be shown that the residual stress tensor $\boldsymbol\tau$ is independent of $\varepsilon$, but we do not provide the details here.

The function $\boldsymbol \Gamma$ is evaluated in the homogeneous state and it is therefore convenient to use the energy function  \eqref{Red-Energy} defined in terms of the principal stretches \eqref{Stretch-homogeneous} and we write
\begin{equation}
\label{Expression-B}
\mathbf{\Gamma}=\phi_1\mathbf e_{\theta}\otimes \mathbf e_{\theta} + \phi_2 \mathbf e_z\otimes \mathbf e_z+ \phi_3\mathbf e_r\otimes \mathbf e_r.
\end{equation}

The reason we introduced the variable $S$ was to identify all terms of $\mathcal{O}(\varepsilon^2)$ in \eqref{Deformation-gradient}. We now return to the original definition $\lambda=\lambda(Z)$ and for clarity of presentation use the shorthand notations  $u=\varepsilon^2\tilde{u}, w=\varepsilon \tilde{w}$.  The energy functional \eqref{Energy} can then be written in the simplified form
\begin{equation}
\label{new-energy}
\mathcal{E}=2\pi\int_{-L}^L \left[G(\lambda)+\mathcal{E}_2\right]\mathrm{d}Z,
\end{equation}
where 
\begin{eqnarray}
G(\lambda)&=&\int_0^A\phi\, R\mathrm{d}R-\frac{1}{2}A^2\lambda \hat{N},\nonumber\\
\mathcal{E}_2&=&\int_0^A\bigg[\phi_1\frac{u}{R}+(\phi_2-\hat{N})w_{Z}+\phi_3u_{R}\nonumber\\&+&\frac{1}{2}\left(\mathcal{A}_{2323}w_{R}^2-\mathcal{A}_{2332}\lambda'\lambda^{-3/2}Rw_{R}+\frac{1}{4}\mathcal{A}_{3232}\lambda'^2\lambda^{-3}R^2\right)\bigg]R\mathrm{d}R,\label{Components}
\end{eqnarray}
where here and in what follows $\lambda'=\mathrm{d}\lambda(Z)/\mathrm{d}Z$.  
Next, we calculate the incompressibility constraint \eqref{Incompressibility} up to $\mathcal{O}(\varepsilon^2)$ and obtain
\begin{equation}
\label{Incom}
2\lambda^{5/2}(Ru)_{R}+R(2\lambda w_{Z}+R\lambda'w_{R})=0,
\end{equation}
which allows to eliminate the variable $u$ in $\eqref{Components}_2$. We use the connection 
\begin{eqnarray}
\int_0^A\left(\phi_1u+\phi_3Ru_{R}\right)\mathrm{d}R&=&\int_0^A\left[\bar{T}_{11}u+\bar{T}_{33}Ru_{R}+\bar{p}\lambda^{1/2}\left(u+Ru_{R}\right)\right]\mathrm{d}R\nonumber \\
&=&\bar{T}_{33}Ru\Big |_0^A+\int_0^A\left(\bar{p}\lambda^{1/2}(Ru)_{R}\right)\mathrm{d}R\nonumber\\
&=&-\int_0^A\bar{p}\left(\frac{w_{Z}}{\lambda}+\frac{R\lambda'w_{R}}{2\lambda^2}\right)R\mathrm{d}R,\label{Replacement-u}
\end{eqnarray}
to express $\eqref{Components}_2$ in the form 
\begin{eqnarray}
\mathcal{E}_2&=&\int_0^A\left[-\bar{p}\left(\frac{w_{Z}}{\lambda}+\frac{R\lambda'w_{R}}{2\lambda^2}\right)R+R(\phi_2-\hat{N})w_{Z}\right.\nonumber\\
&+&\left.\frac{1}{2}R\left(\mathcal{A}_{2323}w_{R}^2-\mathcal{A}_{2332}\lambda'\lambda^{-3/2}Rw_{R}+\frac{1}{4}\mathcal{A}_{3232}\lambda'^2\lambda^{-3}R^2\right)\right]\mathrm{d}R.\label{newlabel}
\end{eqnarray}
In the formulation  \eqref{newlabel}, the term containing the variable $w_Z$  can be eliminated by integration by parts and by using the definition $\bar{T}_{22}=\phi_2-\lambda^{-1}\bar{p}$. This allows to write the energy functional \eqref{new-energy} in the alternative form
\begin{equation}
\mathcal{E}=2\pi\int_{-L}^L \left[G(\lambda)+\int_0^A\mathcal{M}(R,w,w_{R})\mathrm{d}R\right]\mathrm{d}Z+\left[\int_0^A2\pi R\left(\bar{T}_{22}-\hat{N}\right)w\mathrm{d} R\right]\Bigg|_{Z=-L}^{Z=L},
\label{New-energy}
\end{equation}
where 
\begin{equation}
\mathcal{M}(R,w,w_{R})=\frac{1}{2}R\mathcal{A}_{2323}w_{R}^2-\frac{1}{2}R^2\lambda'\left(\mathcal{A}_{2332}\lambda^{-3/2}+\bar{p}\lambda^{-2}\right)w_{R}-\bar{T}_{22,\lambda}\lambda'wR+\frac{1}{8}\mathcal{A}_{3232}\lambda'^2\lambda^{-3}R^3,
\label{M-expr}
\end{equation}
where we introduced the shorthand notation $\bar T_{22,\lambda}=\partial \bar T_{22}/\partial \lambda$.

To make further progress we  assign a specific value to $\lambda$, which allows to obtain the Euler-Lagrange equation and to determine the optimal correction $w$.  Once $w$ is known, it can be eliminated from the energy functional  $\mathcal{E}$ resulting in a one-dimensional theory in terms of $\lambda$.

The Euler-Lagrange equation of $w$ and the required boundary condition are obtained as
\begin{eqnarray}
&&\frac{\mathrm{d}}{\mathrm{d} R}\left[R\mathcal{A}_{2323}w_{R}-\frac{1}{2}R^2\lambda'\left(\mathcal{A}_{2332}\lambda^{-3/2}+\bar{p}\lambda^{-2}\right)\right]=-\bar{T}_{22,\lambda}\lambda'R,\quad 0<R<A,
\label{Equation-W}\\
&&\frac{1}{2}R^2\lambda'\left(\mathcal{A}_{2332}\lambda^{-3/2}+\bar{p}\lambda^{-2}\right)=R\mathcal{A}_{2323}w_{R}\quad\mathrm{on}~ R=0,A.
\label{BounCon}
\end{eqnarray}
The validity of \eqref{BounCon} on $R=0$ is obvious as both sides involve a factor $R$. From \eqref{BounCon} it follows that the inhomogeneous term $-\bar{T}_{22,\lambda}\lambda'R$ in \eqref{Equation-W} must satisfy the  solvability condition
\begin{equation}
\int_0^A\bar{T}_{22,\lambda}\lambda'R\mathrm{d}R=0.
\label{Sol-con}
\end{equation}

Integrating \eqref{Equation-W} with respect to $R$ results in 
\begin{equation}
\label{WR}
w_{R}=g(\lambda,R)\lambda'(Z),
\end{equation}
where
\begin{equation}
\label{glambdaR}
g(\lambda,R)=\frac{1}{2\mathcal{A}_{2323}}R\left(\mathcal{A}_{2332}\lambda^{-3/2}+\bar{p}\,\lambda^{-2}\right)-\frac{1}{R\mathcal{A}_{2323}}\int_0^R\bar{T}_{22,\lambda}t\mathrm{d}t.
\end{equation}
Integration of \eqref{WR} with respect to $R$ gives  the optimal correction term 
\begin{equation}
w=\lambda'(Z) \int_0^Rg(\lambda,t)\mathrm{d}t.
\label{W-sol}
\end{equation}
In the above formula, we have omitted the function arising from integration as it can be absorbed by $\lambda(Z)$.

As anticipated, we now eliminate the dependence of the energy $\mathcal{E}$ on $w$. This is accomplished by using \eqref{WR} and \eqref{W-sol} in \eqref{M-expr} and by integration of $\mathcal{M}(R,w,w_{,R})$ from $0$ to $A$. It follows that
\begin{equation}
\int_0^A\mathcal{M}(R,w,w_R)\mathrm{d}R=(\mathcal{Q}_1+\mathcal{Q}_2+\mathcal{Q}_3)\lambda'(Z)^2,
\label{Int-M}
\end{equation}
where
\begin{eqnarray} 
\mathcal{Q}_1&=&\int_0^A\left[\frac{1}{2}R\mathcal{A}_{2323}g^2(\lambda,R)-\frac{1}{2}R^2\left(\mathcal{A}_{2332}\lambda^{-3/2}+\bar{p}\lambda^{-2}\right)g(\lambda,R)\right]\mathrm{d}R,
\label{Q1}\\
\mathcal{Q}_2&=&-\int_0^A\left(\bar{T}_{22,\lambda}R\int_0^Rg(\lambda,t)\mathrm{d}t\right)\mathrm{d}R=-\int_0^A\left(g(\lambda,R)\int_R^A\bar{T}_{22,\lambda}t\mathrm{d}t\right)\mathrm{d}R,
\label{Q2}\\
\mathcal{Q}_3&=&\int_0^A\frac{1}{8}\mathcal{A}_{3232}\lambda^{-3}R^3 \mathrm{d}R.
\label{Q3}
\end{eqnarray}

Using \eqref{Equation-W}, \eqref{Sol-con} and \eqref{WR} leads to
\begin{equation}
\int_R^A\bar{T}_{22,\lambda}t\mathrm{d}t=-\int_0^R\bar{T}_{22,\lambda}t\mathrm{d}t=R\mathcal{A}_{2323}g(\lambda,R)-\frac{1}{2}R^2\left(\mathcal{A}_{2332}\lambda^{-3/2}+\bar{p}\lambda^{-2}\right),
\label{New-int}
\end{equation}
which is then used to rewrite \eqref{Q2} as 
\begin{equation}
\mathcal{Q}_2=-\int_0^A\left[R\mathcal{A}_{2323}g^2(\lambda,R)-\frac{1}{2}R^2\left(\mathcal{A}_{2332}\lambda^{-3/2}+\bar{p}\lambda^{-2}\right)g(\lambda,R)\right]\mathrm{d}R.
\label{FQ2}
\end{equation}

Expression \eqref{Int-M} is now substituted into \eqref{New-energy} to yield a one-dimensional form of the total potential energy of a hyperelastic, residually stressed cylindrical solid. It has the form
\begin{equation}
\mathcal{E}_\mathrm{1d}=2\pi\int_{-L}^L \left(G(\lambda)+\frac{1}{2}D(\lambda)\left(\lambda'\right)^2\right)\mathrm{d}Z+C(\lambda)\lambda'\Big|_{-L}^{L},
\label{1d-energy}
\end{equation}
where we recall that $\lambda=\lambda(Z)$ and $\lambda'=\mathrm{d}\lambda(Z)/\mathrm{d}Z$. In \eqref{1d-energy} we introduced the functions $D(\lambda), C(\lambda)$, which have the explicit forms
\begin{eqnarray}
D(\lambda)&=&\int_0^A\left(\frac{1}{4}\mathcal{A}_{3232}\lambda^{-3}R^3-R\mathcal{A}_{2323}g^2(\lambda,R)\right)\mathrm{d}R,
\label{Exp-D}\\
C(\lambda)&=&2\pi\int_0^Ag(\lambda,R)\int_R^A\left(\bar{T}_{22}-\hat{N}\right)t\mathrm{d}t.
\label{Exp-c}
\end{eqnarray}
It can be checked that without residual stress \eqref{Exp-D} will recover Equation (2.29b) in \cite{Audoly2016}.

The required equilibrium equation that governs the axial stretch $\lambda$ is obtained by minimizing the energy \eqref{1d-energy} with respect to $\lambda$. The results is the second-order nonlinear differential equation 
\begin{equation}
G'(\lambda)-\frac{1}{2}D'(\lambda)\left(\lambda'\right)^2-D(\lambda)\lambda''=0,\label{1d-model}
\end{equation}
where $G'(\lambda)=\mathrm d G/\mathrm d\lambda$ and $D'(\lambda)=\mathrm d D/\mathrm d\lambda$.  It can be seen that $Z$ does not explicitly appear in the integrand of \eqref{1d-energy}, instead the integrand depends on $Z$ through $\lambda$. This is consistent with the  translational invariance  in $Z$ of the current problem. According to the Beltrami identity, equation \eqref{1d-model} admits a first integral 
\begin{equation}
G(\lambda)-\frac{1}{2}D(\lambda)\left(\lambda'\right)^2=\textrm{constant}.\label{FirstInt}
\end{equation}

Once $\lambda$ is determined , the profile of the localized bifurcation can be characterized according to \eqref{Deformation}. When compared to the solution  without residual stress,  the coefficients here  depend on $R$. Furthermore, the explicit expression of $D(\lambda)$ requires integration, which may be difficult to obtain for some material models. Here we use the numerical integration scheme combined with the symbolic software package Mathematica \cite{Wolfram2019} to obtain solutions.

\section{Material model and bifurcation point} \label{Material model}

The invariants based constitutive theory \cite{Hoger1993}, using the right Cauchy-Green tensor $\mathbf C=\mathbf F^{\mathrm T}\mathbf F$ and the residual stress field (9) as the independent variables,  leads to a transversely isotropic response. Accordingly, we consider the energy as a scalar valued function in terms of the invariants $I_1$ and $I_5$. Specifically, let the mechanical properties be given by an incompressible Gent model with residual stress as follows
\begin{equation}
\label{GentModel}
W=-\frac{1}{2}\mu J_\mathrm{m}\ln\left(1-\frac{I_1-3}{J_\mathrm{m}}\right)+\frac{1}{2}\left(I_5-\tr {\boldsymbol \tau}\right)+\frac{1}{4}\kappa\left(I_5 - \tr {\boldsymbol \tau} \right)^2,
\end{equation}
where $\mu>0$ is the shear modulus in the reference configuration, $I_1=\operatorname{tr}\left(\mathbf{F}^\mathrm{T}\mathbf{F}\right)$, $I_5=\operatorname{tr}\left(\boldsymbol\tau\mathbf{F}^\mathrm{T}\mathbf{F}\right)$, $J_\mathrm{m}$ is a  parameter related to the the maximum material extensibility, and $\kappa$ is a non-negative  constant. We mention that for $J_\mathrm{m}\rightarrow\infty$ the Gent model \eqref{GentModel}  reduces to the neo-Hookean model, which we used in \cite{Liu2023MMS}.  There, we found that for increasing load localized bifurcation with zero mode number  occurs first. 

For the incremental deformation and  the residual stress   \eqref{Tau-exp}, the principal invariants  can be calculated as 
\begin{equation}\label{Invariants}
\begin{aligned}
I_1~~~&=~~~2\lambda^{-1}+\lambda^2+\dfrac{2u}{\sqrt {\lambda}R}+2\lambda w_{Z}+\dfrac{R^2 \lambda_{Z}^2}{4 \lambda^3}+ w^2_{R}+2\dfrac{u_{R}}{\sqrt {\lambda}}
,\\
I_5~~~&=~~~\tau_{\Theta\Theta}\left(\dfrac{1}{\lambda}+\dfrac{2u}{\sqrt {\lambda}R}\right)+\tau_{RR}\left(\dfrac{1}{\lambda}+w^2_{R}+\dfrac{2u_{R}}{\sqrt {\lambda}}\right).
\end{aligned}
\end{equation}

To obtain an explicit form of \eqref{Tau-exp}, we take
\begin{equation}
\tau_{RR}=\nu(R^2-A^2),
\label{Tau33}
\end{equation}
where the parameter $\nu$ specifies the magnitude and the sign of the residual stress. It then follows from the equilibrium equation \eqref{Eq-residual-stress} that 
\begin{equation}
\tau_{\Theta\Theta}=\nu(3R^2-A^2).
\label{Tau11}
\end{equation}
For the homogeneous deformation,  the forms of the Lagrange multiplier and of the axial  force  $N$  can  be determined as
\begin{eqnarray}
\bar{p}&=&\frac{\mu  J_\mathrm{m}}{\lambda  J_\mathrm{m}-\lambda ^3+3 \lambda -2}-\frac{2 \kappa  (\lambda -1) \nu ^2 \left(A^2-R^2\right)^2}{\lambda ^2},\label{pressure}\\
N&=&-\frac{\pi  A^2 (\lambda -1) \Big(\lambda  J_\mathrm{m} \left(2 A^4 \kappa  \nu ^2+3 \lambda\mu\left(\lambda^2 +3 \lambda +1\right)\right) -2 A^4 \kappa \left(\lambda ^3-3 \lambda +2\right) \nu ^2\Big)}{3 \lambda ^3 \left(-\lambda  J_\mathrm{m}+\lambda ^3-3 \lambda +2\right)},
\label{force}
\end{eqnarray}
and localized necking or bulging occurs for  
\begin{equation}
\label{Dforce-lambda}
\frac{\mathrm{d} N}{\mathrm{d}\lambda}=0,
\end{equation}
regardless of the specific loading sequence \cite{Liu2023MMS}. 

The nominal stress component $\bar{T}_{22}$ is needed to evaluate \eqref{glambdaR} and takes the form 
\begin{equation}
\bar{T}_{22}=-\frac{2 \kappa  (\lambda -1) \nu ^2 \left(A^2-R^2\right)^2 \left(\lambda ^3-\lambda  J_\mathrm{m}-3 \lambda +2\right)+(\lambda ^3 +1)\lambda ^2 \mu 
   J_\mathrm{m}}{\lambda ^3 \left(\lambda ^3-\lambda  J_\mathrm{m}-3 \lambda +2\right)}.\label{T22}
\end{equation}
The function $G(\lambda)$ is given by
\begin{eqnarray}
\notag G(\lambda)&=&\frac{A^6 \kappa  \nu ^2}{6 \lambda ^2}-\frac{1}{4} A^2 \mu  J_m \ln \left(1-\dfrac{\lambda ^2+2 \lambda^{-1}-3}{J_m}\right)-\dfrac{A^6 \kappa  \nu ^2}{3 \lambda }+\frac{1}{6}A^6 \kappa  \nu ^2\\
&+&\dfrac{A^2 \lambda  \left(\lambda _{\infty }-1\right) \left(J_m\lambda _{\infty } \left(2 A^4 \kappa  \nu ^2+3 \mu  \lambda _{\infty }^3+3 \mu \lambda _{\infty }^2+3 \mu  \lambda _{\infty }\right)-2 A^4 \kappa  \nu ^2\left(\lambda _{\infty }^3-3 \lambda _{\infty }+2\right)\right)}{6 \lambda _{\infty}^3 \left(-J_m \lambda _{\infty }+\lambda _{\infty }^3-3 \lambda _{\infty}+2\right)},\label{G-expression}
\end{eqnarray}
where  the stretch at infinity is denoted by $\lambda _{\infty}$.

For the Gent material with residual stress we find that 
\begin{eqnarray}
\notag\mathcal{A}_{2323}&=&\nu\left(R^2-A^2\right)\left(1+\frac{2 \kappa  (\lambda -1) \nu \left(A^2-2R^2\right)}{\lambda }\right)+\frac{\mu \lambda J_\mathrm{m}}{\lambda  J_\mathrm{m}-\lambda ^3+3 \lambda -2},\\
\mathcal{A}_{3232}&=&\frac{\mu \lambda J_\mathrm{m}}{\lambda  J_\mathrm{m}-\lambda ^3+3 \lambda -2},\label{Moduli}\\\notag
\mathcal{A}_{2332}&=&0.
\end{eqnarray}

Before proceeding further, we introduce the following dimensionless variables 
\begin{equation}
w^*=\frac{w}{A},\quad u^*=\frac{u}{A},\quad L^*=\frac{L}{A},\quad W^*=\frac{W}{\mu},\quad N^*=\frac{N}{\mu A^2},\quad a^*=\frac{a}{A},\quad\kappa^*=\kappa\mu,\quad\nu^*=\frac{\nu A^2}{\mu}.
\end{equation}
In the following, for clarity of presentation, we use these dimensionless measures without the superposed asterisk and for illustrative purpose we take $L/A=10$.

A linear bifurcation analysis is used in \cite{Liu2023MMS} to determine the conditions for localized necking or bulging of a residually stressed neo-Hookean circular cylindrical solid. In particular,  Figure 3 in \cite{Liu2023MMS} depicts the critical stretch $\lambda_{\mathrm {cr}}$ as a function of $\kappa \nu^2$ and identifies  $\kappa \nu^2=29.5585$ as the lowest value when localized bifurcation occurs at  $\lambda_{\mathrm {cr}}=\lambda_{\mathrm m}=2.1069$. When the constant stretch $\lambda< \lambda_{\mathrm {m}}$, a gradual increase in residual stress initiates localized necking. However when $\lambda > \lambda_{\mathrm {m}}$ the critical value of the residual stress generates localized bulging, see Figure 5 in \cite{Liu2023MMS}.

\begin{figure}[!ht]
        \centering
	\includegraphics[scale=0.8]{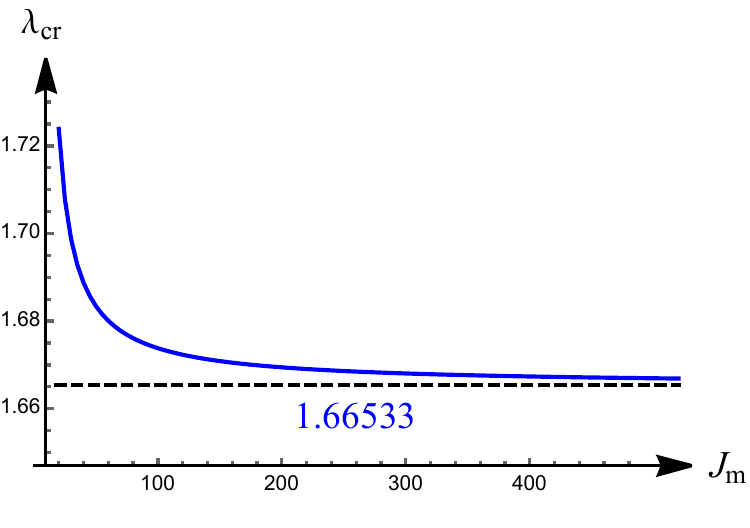}
	\caption{The graph shows $\lambda_\textrm{cr}$ as a function of the material extensibility  $J_\textrm{m}$ for $\kappa\nu^2=50$. The dashed line corresponds to the asymptote as $J_\textrm{m}\rightarrow\infty$, when the critical stretch approaches $\lambda_\textrm{cr}=1.66533$ of the neo-Hookean material.}\label{lambdacr}
\end{figure}

To verify the current formulation, we use a representative value $\kappa\nu^2=50$ and solve the bifurcation condition \eqref{Dforce-lambda} for different values of $J_{\mathrm m}$ and compare results for $J_{\mathrm m} \rightarrow \infty$ with those reported in  \cite{Liu2023MMS}.   Figure \ref{lambdacr}  shows  that the critical stretch $\lambda_\textrm{cr}$ is a monotonically decreasing function of $J_\textrm{m}$, indicating that a lower material extensibility limit  delays the initiation of localized instability. As $J_\textrm{m}\rightarrow\infty$, the Gent model  reduces to the neo-Hookean model and the critical stretch $\lambda_{\mathrm {cr}}\approx 1.66533$, which coincides with the results shown in Figure 2 in  \cite{Liu2023MMS}.

\begin{figure}[!ht]
	\centering\includegraphics[scale=0.8]{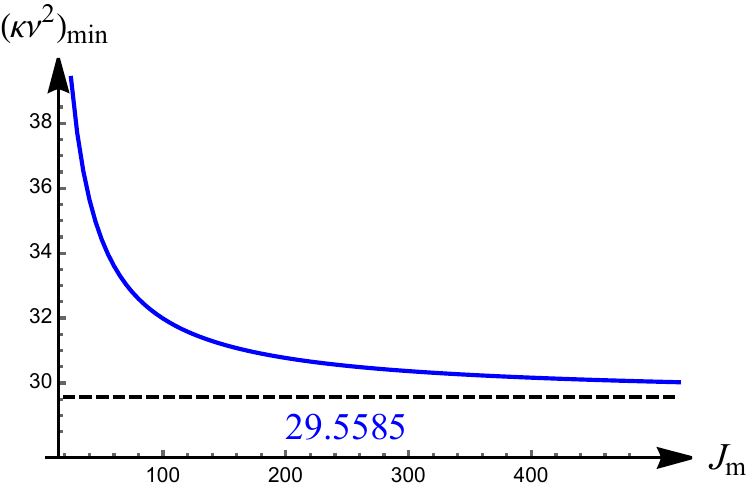}
	\caption{This figure shows  $\left(\kappa \nu^2\right)_{\mathrm{min}}$ as a decreasing function of $J_{\mathrm m}$ and in the limit, as $J_{\mathrm m}\rightarrow\infty$, it approaches  $\kappa \nu^2=29.5585$ of the neo-Hookean model.}\label{kappanu2min-jm}
\end{figure}

For the Gent material \eqref{GentModel},  the conditions $\textrm{d}N/\textrm{d}\lambda=0$ and $\textrm{d}^2N/\textrm{d}\lambda^2=0$ determine  the minimum values of $\kappa \nu^2$ for localized bifurcation as a function of $J_{\mathrm m}$.  Figure \ref{kappanu2min-jm} shows that $\left(\kappa \nu^2\right)_{\mathrm{min}}$ is a decreasing function of $J_{\mathrm m}$ and as $J_{\mathrm m}\rightarrow\infty$ approaches  $\kappa \nu^2=29.5585$   of the neo-Hookean material. The graph also shows that there exists a  critical value of $J_{\mathrm m}=J_{\mathrm m}^*$ when $\left(\kappa \nu^2\right)_{\mathrm{min}} \rightarrow \infty$. To determine this value we take the limit of $\kappa\nu^2\rightarrow\infty$ in the bifurcation condition \eqref{Dforce-lambda} and obtain $\lambda=1.5$, which is then substitute back into $\textrm{d}N/\textrm{d}\lambda=0$ resulting in $J_\textrm{m}^*=0.58333$. For $J_\textrm{m}<J_{\mathrm m}^*$ localized bifurcation does not occur regardless the amount of the residual stress. This is similar to the options mentioned in \cite{Liu2019} on how to prevent localized bulging in inflated tubes. 

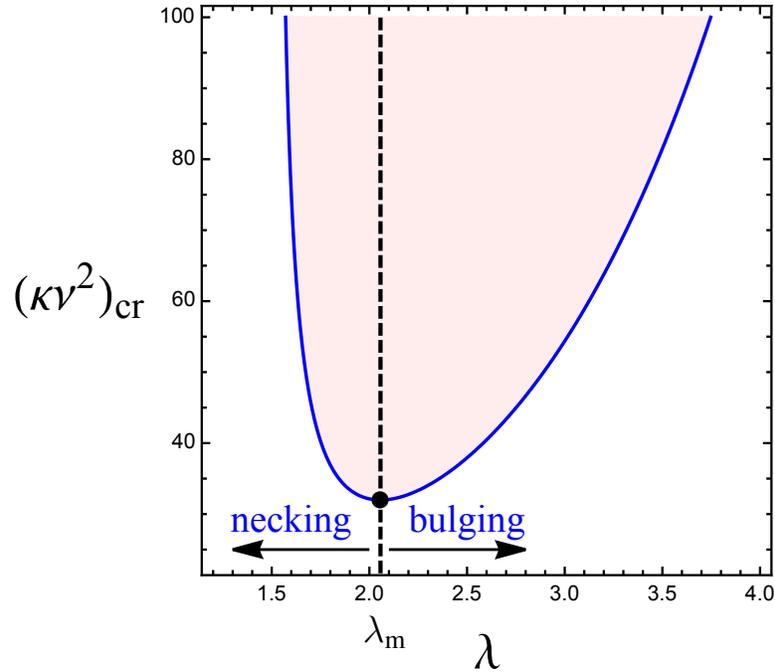
\begin{figure}[htb!]
\begin{picture}(0,0)(0,0)
\put(135,16){\Large{$\lambda_{\textrm{m}}$}}
\end{picture}
	\centering\includegraphics[scale=0.8]{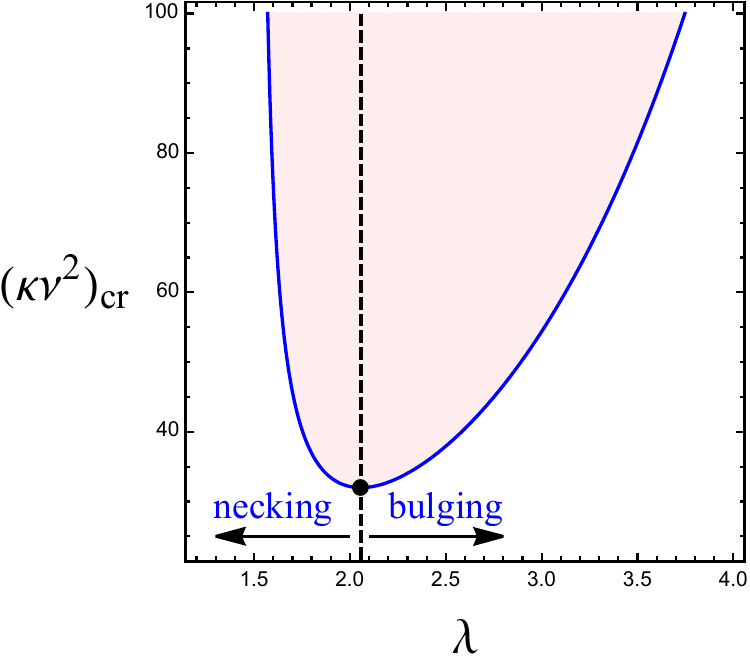}
	\caption{The relation between the critical residual stress $(\kappa\nu^2)_\textrm{cr}$ and the pre-stretch $\lambda$ for $J_{\mathrm m}=100$. The black dot represents the lowest value $\kappa \nu^2=31.9855$ when the axial pre-stretch   $\lambda_{\mathrm {m}}=2.0567$. An increase in residual stress with constant stretch $\lambda<\lambda_{\mathrm {m}}$ initiates localized necking, with $\lambda>\lambda_{\mathrm {m}}$  localized bulging.}\label{bifurcation-curve}
\end{figure}

In Figure  \ref{bifurcation-curve} we present the dependence of the critical residual stress field $\left(\kappa \nu^2\right)_{\mathrm {cr}}$ on the amount of pre-stretch $\lambda$ for $J_{\mathrm m}=100$ and indicate  $\kappa \nu^2=31.9855$ as the lowest value when localized bifurcation occurs at  $\lambda_{\mathrm {m}}=2.0567$. Consistent with the results in \cite{Liu2023MMS} , an increase in residual stress for constant stretch $\lambda<\lambda_{\mathrm {m}}$ initiates localized necking, for $\lambda>\lambda_{\mathrm {m}}$  localized bulging. 

\begin{figure}[htb!]
	\centering\includegraphics[scale=0.8]{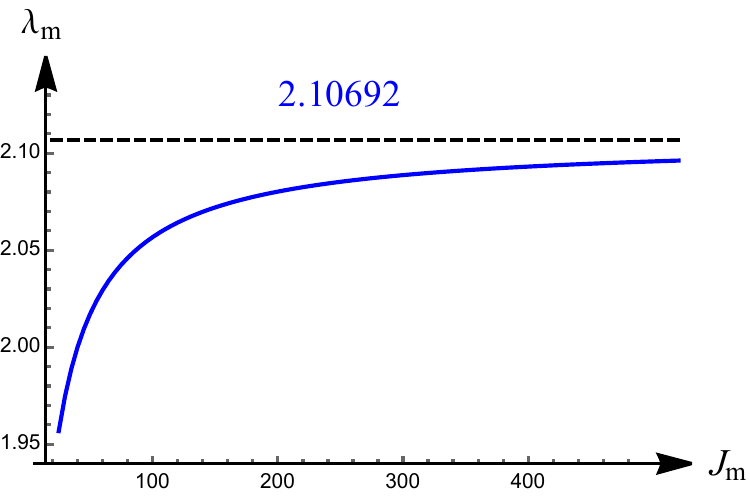}	
	\caption{The transition stretch $\lambda_{\mathrm {m}}$ as a function of $J_{\mathrm {m}}$ for constant $\kappa \nu^2=31.9855$. When the material extensibility  limit $J_{\mathrm m}\rightarrow\infty$, the  stretch $\lambda_{\mathrm {m}}$ approaches the value of the neo-Hookean model $\lambda_{\mathrm {m}}=2.1069$. }
\label{lambdam-jm}
\end{figure}

In Figure \ref{lambdam-jm} we keep  $\kappa \nu^2=31.9855$ constant to obtain $\lambda_{\mathrm {m}}$ as a function of $J_{\mathrm {m}}$. When the material extensibility  limit $J_{\mathrm m}\rightarrow\infty$ the transition stretch $\lambda_{\mathrm {m}}$ approaches the value of the neo-Hookean model $\lambda_{\mathrm {m}}=2.1069$.

\section{Nonlinear one-dimensional model} \label{Nonlinear}

We now focus our attention on using the one-dimensional  model  \eqref{1d-model} to obtain a better understanding of the initiation and growth of a local bifurcation.  First, we substitute \eqref{pressure}, \eqref{force}, \eqref{T22}, and \eqref{Moduli} into \eqref{glambdaR} and \eqref{Exp-D} and use the results together with \eqref{G-expression} to obtain the explicit form of \eqref{1d-model}.  We use the centered finite difference method to solve  the resulting nonlinear differential equation  subject to  well-posed boundary conditions. We discretize the half-length $[0,L]$ of the solid   cylinder into $k$ intervals of step size $h=L/k$, determine the axial coordinate of node $j$ by $Z_j=j h$ and denote the corresponding stretch   $\lambda\left(Z_j \right)$ by $\lambda_j$. The solution of the one-dimensional model is then reduced to the solution of a set of algebraic equations of the form 
\begin{equation}
G'(\lambda_j)-\frac{1}{2}D'(\lambda_j)\left(\frac{\lambda_{j+1}-\lambda_{j-1}}{2h}\right)^2-D(\lambda_j)\left(\frac{\lambda_{j+1}-2\lambda_j+\lambda_{j-1}}{h^2}\right)=0,\quad j=1,2,\cdots,k-1.
\label{Model-discretized}
\end{equation}

Suitable boundary conditions will be specified for the two loading conditions considered. In the first, the residual stress  parameter $\nu$ is maintained constant, while the axial stretch is increased until a critical value is reached. In \cite{Liu2023MMS} we found that with increasing stretch and with $\kappa\nu^2$ larger than a critical value,  localized necking occurs. In the second case, we keep the pre-stretch  constant and monotonically increase the residual stress  $\nu$ until bifurcation occurs. Figure \ref{bifurcation-curve} in this paper and Figure 5 in \cite{Liu2023MMS}  show that the bifurcation  changes character from necking to bulging as the pre-stretch becomes larger than a critical value.

In  \cite{Yu2023} it is shown that the one-dimensional model accurately describes the entire evolution process of bulging or necking for an  inflated circular cylindrical tube of arbitrary wall thickness.  In particular, the numerical results of a nonlinear finite element analysis are used to validate and verify the accurateness of the  model.

\subsection{Fixed residual stress}\label{Fixed-stress}

In this loading sequence the residual stress parameter $\nu$ is constant while  the axial stretch is taken as the loading parameter. The one-dimensional model \eqref{1d-model} only admits a localized solution if $D(\lambda)>0$. When integrating \eqref{Exp-D}  from 0 to $A$ we recognize that  at the location
\begin{equation}
R_0=\left(\frac{\lambda+6 \kappa  \lambda  \nu -6\kappa  \nu -
  \left(\left(\lambda  \left(2 \kappa  \nu -1\right)-2 \kappa  \nu
   \right)^2-\dfrac{\lambda^216 \kappa  (\lambda -1) J_m}{-\lambda  J_m+\lambda
   ^3-3 \lambda +2}\right)^{1/2}}{8\kappa  (\lambda -1) \nu }\right)^{1/2},
\end{equation} 
the modulus  $\mathcal{A}_{2323}=0$, which  generates a singularity in \eqref{glambdaR}.  In the linear  analysis \cite{Liu2023MMS}, a similar singularity  was classified as a regular singular point.  Here, we use the same strategy and take the Cauchy principal value of the improper integral in the numerical integration scheme.  We write 
\begin{equation}
D(\lambda)=\int_0^{R_0-\delta}\left[\frac{1}{4}\mathcal{A}_{3232}\lambda^{-3}R^3-R\mathcal{A}_{2323}g^2(\lambda,R)\right]\mathrm{d}R+ \int_{R_0+\delta}^A\left[\frac{1}{4}\mathcal{A}_{3232}\lambda^{-3}R^3-R\mathcal{A}_{2323}g^2(\lambda,R)\right]\mathrm{d}R
\label{D-principal}
\end{equation}
where $\delta$ is a small constant, here taken as  $\delta=10^{-6}$.

\begin{figure}[!ht]
	\centering\includegraphics[scale=0.8]{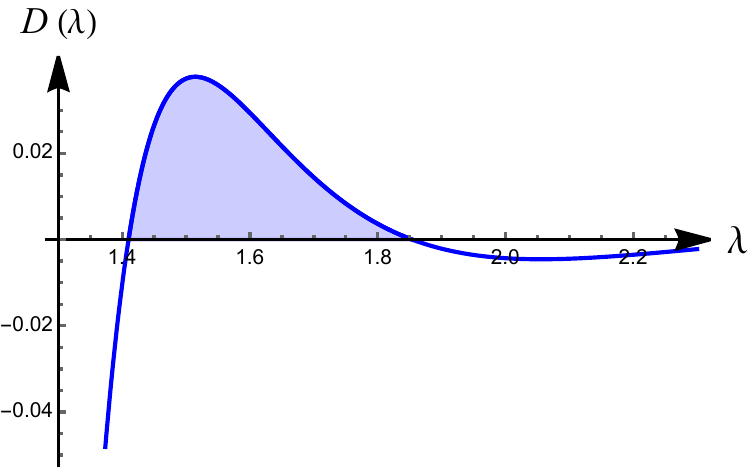}
	\caption{The dependence of $D(\lambda)$ on $\lambda$ when $J_\textrm{m}=100,\kappa=0.5$ and  $\nu=10$ with roots located at  $\lambda_\textrm{t1}=1.40991$ and $\lambda_\textrm{t2}=1.85396$.}
	\label{Dlambda}
\end{figure}

To investigate if \eqref{1d-model} admits a localized solution, we  compute the changes of $D(\lambda)$ as a function $\lambda$ using   $J_\textrm{m}=100, \kappa=0.5$ and   $\nu=10$.  Figure \ref{Dlambda} shows that $D(\lambda)$ has two roots in the interval  $1<\lambda<2.2$. Specifically, these are located at  $\lambda_\textrm{t1}=1.40991$ and $\lambda_\textrm{t2}=1.85396$ and compare to the critical stretch $\lambda_\textrm{cr}=1.67381$, where localized bifurcation occurs. The mathematical nature of the current problem is identical to the one obtained by replacing the residual stress by a surface tension to generate localization in inflated tubes. There, it has been shown that the bifurcation nature is subcritical, a condition of bifurcation  we adopt here as well \cite{Fu2008, Emery2021b, Emery2023, Ye2020, Fu2021, Yu2023}.

In infinitely long  cylindrical solids or tubes it is customary to denote the stretch at infinity by $\lambda_\infty$. When localized bifurcation  occurs, the stretch  no longer remains uniform and  becomes a function of $Z$. For long cylindrical solids of  half-length $L$, the exact value of $\lambda_{\mathrm L}$ has an exponentially small difference from  $\lambda_\infty$, a difference we neglect in the following  \cite{Wang2020}. 

Based on the principle of translational invariance, we   assume that localization initiates at $Z=0$. Then, it is convenient to use the difference between $\lambda(0)$ and $\lambda_\infty$ to identify the bifurcation nature. For subcritical bifurcation, the loading parameter $\lambda_\infty$ will be less than $\lambda_\textrm{cr}$ or, equivalently, $a_{\infty}>a_\mathrm{cr}$. According to the Noether's theorem  \cite{Noether1918}, the evolution of $\lambda(0)$ in the post-bifurcation regime can be obtained by solving
\begin{equation}
G(\lambda(0))=G(\lambda_\infty),
\label{amplitude}
\end{equation}
where the form of $G(\lambda)$ is given in \eqref{G-expression}. Note that \eqref{amplitude} can also be obtained from \eqref{FirstInt} by setting the derivative  $\lambda'=0$.

\begin{figure}[htb!]
	\centering\includegraphics[scale=0.8]{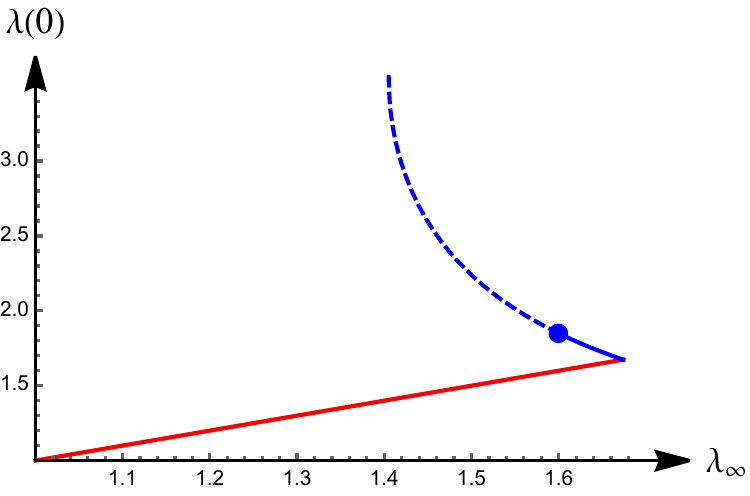}\caption{The bifurcation diagram for $J_\textrm{m}=100$, $\kappa=0.5$ and $\nu=10$.   The homogeneous response $\lambda_\infty=\lambda(0)$ is shown in red, while  the non-trivial solution is depicted in blue.}\label{lambdazero-lambdainf}
\end{figure}

The bifurcation diagram for $J_\textrm{m}=100$, $\kappa=0.5$ and $\nu=10$ is obtained from  \eqref{amplitude} and is shown in  Figure \ref{lambdazero-lambdainf}.  The homogeneous response $\lambda_\infty=\lambda(0)$ is shown in red, while  the non-trivial solution is depicted in blue.  Bifurcation occurs when $\lambda_\infty=\lambda(0)=1.67381$ and, subsequently,  $\lambda(0)$ increases with   $\lambda_\infty$ getting smaller. The  deformed radius  given  by $a(Z)=\lambda(Z)^{-1/2}$, indicates localized necking. The blue dot, in particular, corresponds to $\lambda_\infty= 1.59836$  and $\lambda(0)=\lambda_\textrm{t2}=1.85396$, the latter being a root of $D(\lambda)$, see Figure \ref{Dlambda}. A further reduction in $\lambda_\infty$ results in  $D(\lambda)<0$ and model \eqref{1d-model} no longer admits a localized solution. 

\begin{figure}[!ht]
	\centering\includegraphics[scale=0.6]{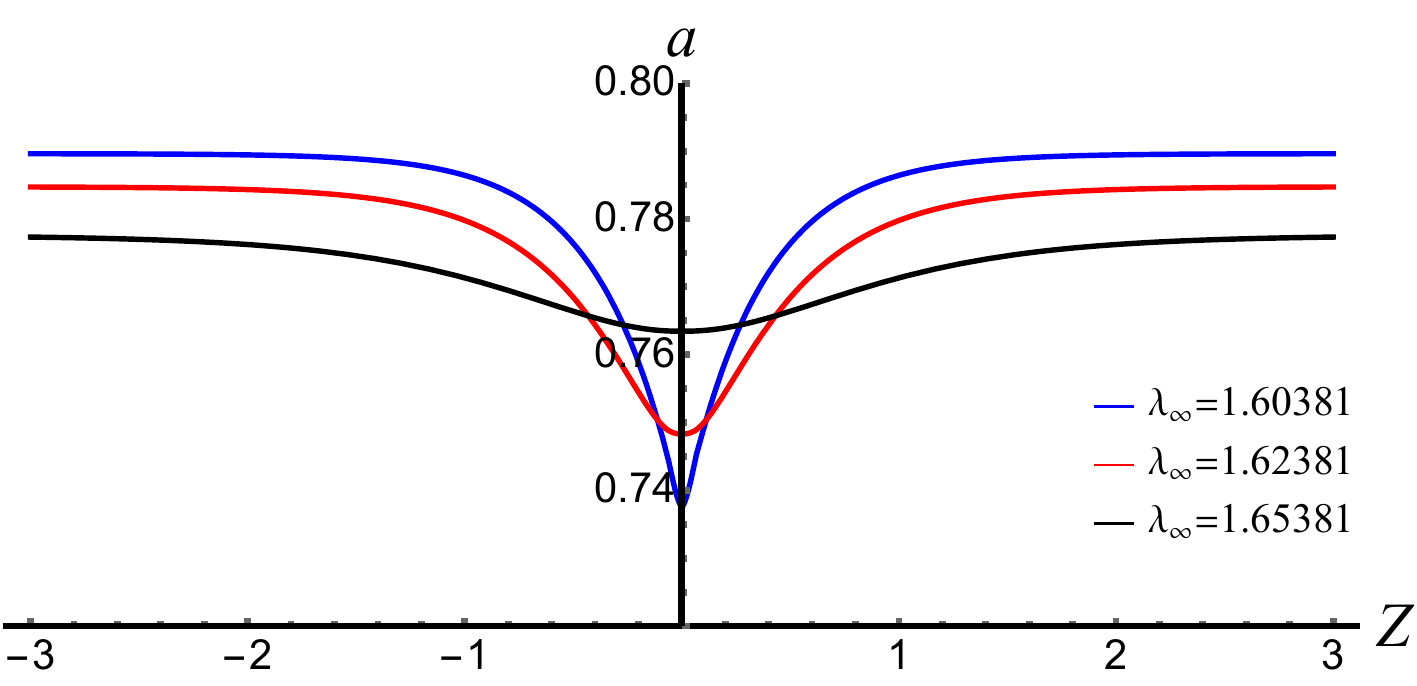}\caption{Dependence of the deformed radius $a$ on $Z$ for various values of $\lambda_\infty$. For clarity of presentation, we show localized necking from $Z=-3$ to $Z=3$.}\label{profile-Jm100-fixednu}
\end{figure}

To obtain the   radius as a function of $Z$  in the post-bifurcation regime, we numerically solve equation \eqref{1d-model} subject to the boundary conditions 
\begin{equation}
\lambda(0)=\lambda_0,\quad \lambda'(0)=0,
\label{1d-bc}
\end{equation}
where  $\lambda_0$ is obtained from \eqref{amplitude} for given $\lambda_\infty$ and $\lambda'(0)=0$ implies symmetry in the deformation. We use the finite difference method with $k=50$ intervals and Newton's method to solve the algebraic equations \eqref{Model-discretized} for  representative values $\lambda_\infty=1.65381, 1.62381$ and $1.60381$. The deformed radius  $a(Z)=\lambda(Z)^{-1/2}$ is shown in Figure \ref{profile-Jm100-fixednu} for  different  $\lambda_\infty$.

\begin{figure}[!ht]
	\centering
	\subfigure[The stretch at the bulged region as a function of $J_\textrm{m}$. ]{\includegraphics[scale=0.6]{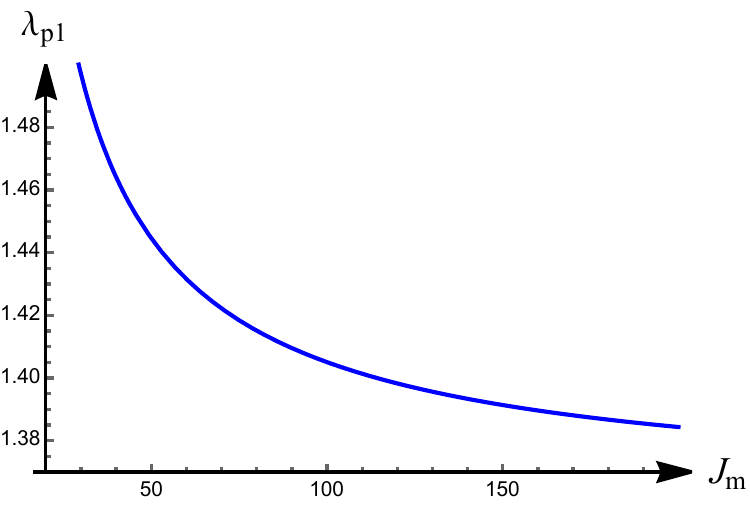}}
	\subfigure[The stretch at the necked region as a function of $J_\textrm{m}$.]{\includegraphics[scale=0.6]{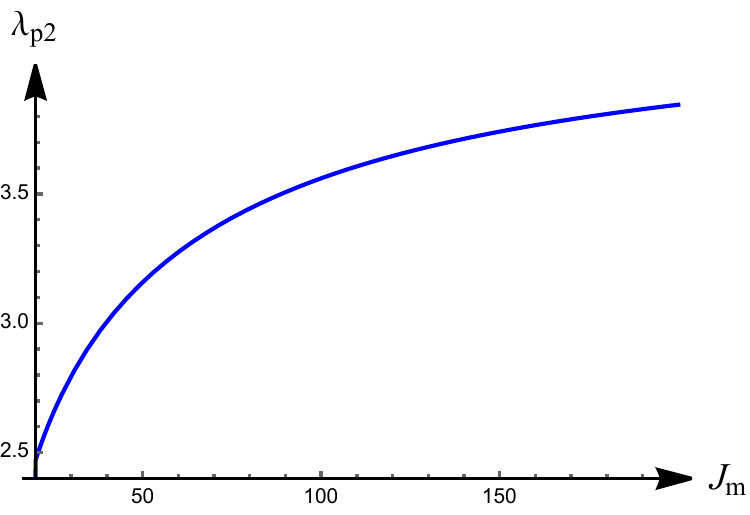}}
	\caption{The dependence of two Maxwell stretches on $J_\textrm{m}$. We understand that in the Maxwell state the deformation is composed of two uniform deformations joined by a transition zone, and we label this by a two-phase deformation as well.}\label{Maxwell-stretch}
\end{figure}

In \cite{Liu2023MMS} we characterized  localized bifurcation, the corresponding  change in radius and the  propagation as  a the two-phase deformation consisting of necked and bulged regions. These are connected by a transition zone, which simply translates along the axial direction. During propagation, the stretches of each phase are unchanged and are determined by  Maxwell's equal-area rule. The change in radius is visualized by the solid/dashed blue curve  in Figure \ref{lambdazero-lambdainf} and is completed once the Maxwell values of stretch are reached, i.e. the end of the blue curve. The  Maxwell values of stretch of the bulged and necked regions  are denoted by $\lambda_\textrm{p1}$, $\lambda_\textrm{p2}$, respectively, and  are shown  as a function of the material extensibility limit $J_{\mathrm m}$ in Figure \ref{Maxwell-stretch}. We find that $\lambda_\textrm{p2}$ is a monotonically increasing function, while $\lambda_\textrm{p1}$ has the opposite behavior.

We also find that the magnitude of the local minimum of  $D(\lambda)$ depends  on $J_\textrm{m}$, see Figure \ref{Dlambda}. For what follows, it is interesting to determine the value  $J_\textrm{m}$ for which the local minimum of $D(\lambda)=0$.  This is given by the solution of
\begin{equation}
\frac{\textrm{d}D(\lambda)}{\textrm{d}\lambda}=0,\quad D(\lambda)=0,
\end{equation}
resulting in  $J^\mathrm{cr}_\mathrm{m}=33.5637$ with the root  increasing to $\lambda_\mathrm{t2}=2.03201$. Both Maxwell stretches now lie in the interval $(\lambda_\textrm{t1},\lambda_\textrm{t2})$ as $J_{\mathrm m} < J_{\mathrm m}^{\mathrm{cr}}$ and  the one-dimensional model will be able to capture the  necking initiation and growth evolution until the two-phase deformation develops, i.e.  $D(\lambda)$ is always positive.

\begin{figure}[!ht]
	\centering\includegraphics[scale=0.8]{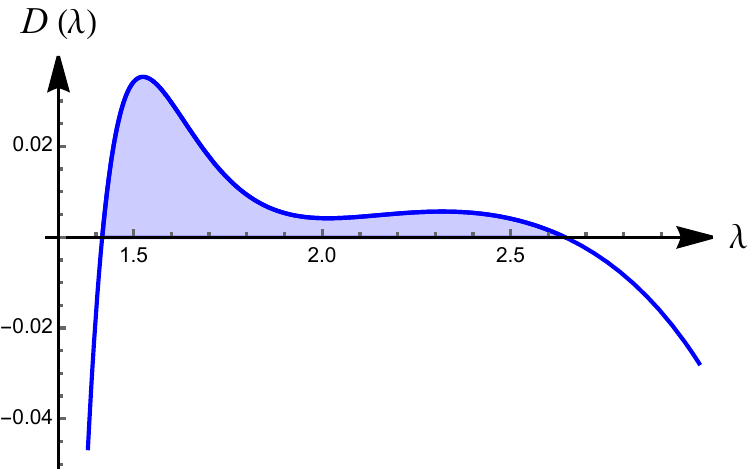}\caption{The dependence of $D(\lambda)$ on $\lambda$ when $J_\textrm{m}=20$, $\kappa=0.5$, and $\nu=10$ with the roots located at $\lambda_\textrm{t1}=1.4164$ and $\lambda_\textrm{t2}=2.64427$.}\label{Dlambda-Jm20}
\end{figure}

\begin{figure}[htb!]
	\centering\includegraphics[scale=0.8]{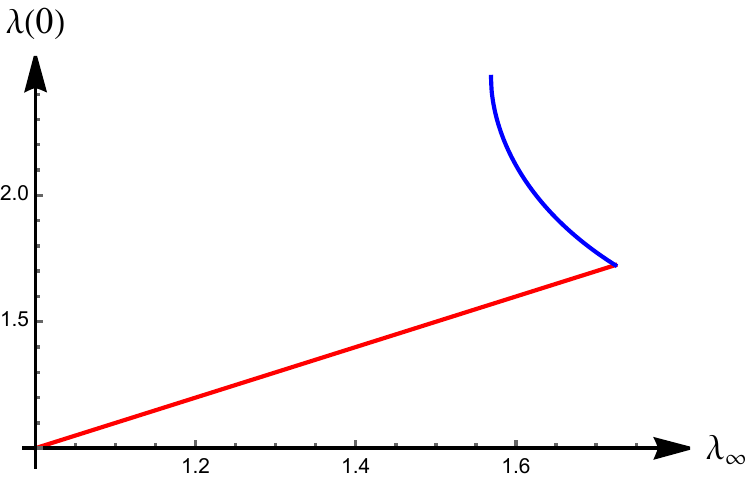}\caption{The bifurcation diagram when $J_\textrm{m}=20$, $\kappa=0.5$, and $\nu=10$. The red line represents  the homogeneous deformation, the blue  depicts the evolution of necking.}\label{lambdazero-lambdainf-Jm20}
\end{figure}

Take $J_\textrm{m}<J^\textrm{cr}_\textrm{m}=20, \kappa=0.5, \nu=10$ and explore the necking evolution  using the model \eqref{1d-model}. Figure \ref{Dlambda-Jm20} illustrates the relation between $D(\lambda)$ and $\lambda$ with the roots now located at   $\lambda_\textrm{t1}=1.4164$ and $\lambda_\textrm{t2}=2.64427$. The corresponding Maxwell values of stretch are calculated as $\lambda_\textrm{p1}=1.56876>\lambda_\textrm{t1}$ and $\lambda_\textrm{p2}=2.46839<\lambda_\textrm{t2}$. The bifurcation diagram is again obtained from  \eqref{amplitude} and is shown in Figure \ref{lambdazero-lambdainf-Jm20}, where  the deformation is homogeneous until  the value  $\lambda_{\infty}=\lambda(0)=\lambda_\textrm{cr}=1.72389$ is reached. The blue curve shows the evolution of the necking amplitude, which terminates at the Maxwell values of stretch.

\begin{figure}[!ht]
	\centering\includegraphics[scale=0.6]{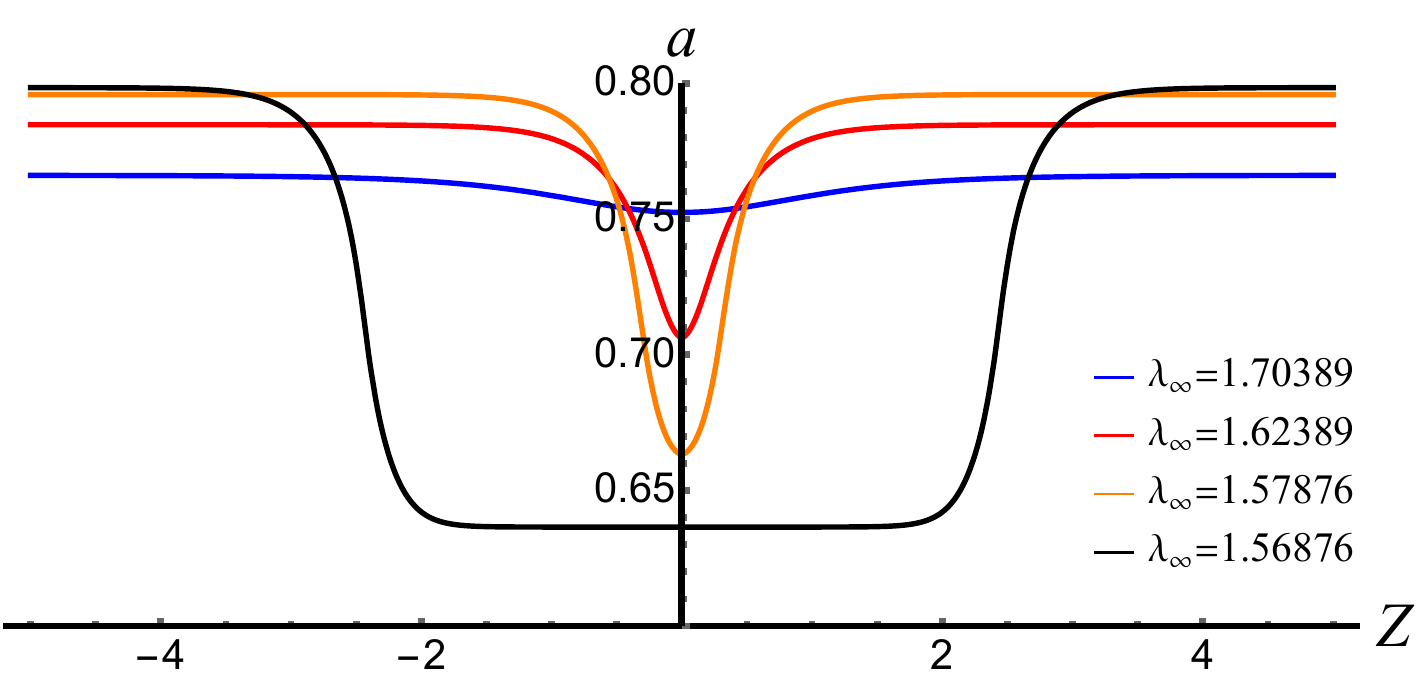}\caption{The evolution of the necking profiles for different values of $\lambda_\infty$ with $J_\textrm{m}=20$, $\kappa=0.5$, and $\nu=10$. For clarity of presentation we restrict attention to the region  $-5\leqslant Z\leqslant5$.}\label{profile-Jm20-fixednu}
\end{figure}

To calculate the axial stretch $\lambda(Z)$ for different values of $\lambda_{\infty}$, we follow the  procedure used to obtain the graphs in Figure \ref{profile-Jm100-fixednu}. Here, we consider $\lambda_{\infty}=1.70389, 1.62389, 1.57876$ and $1.56876$  and show the corresponding  changes of radius $a(Z)=\lambda^{-1/2}(Z)$ in Figure   \ref{profile-Jm20-fixednu}. For the special case where the value of $\lambda_\infty$ is equal to the Maxwell stretch $\lambda_\textrm{p1}=1.56876$, the one-dimensional model  \eqref{1d-model} is capable to simulate  the initiation and the entire growth process.

\begin{figure}[!ht]
	\centering
	\subfigure{\includegraphics[scale=0.6]{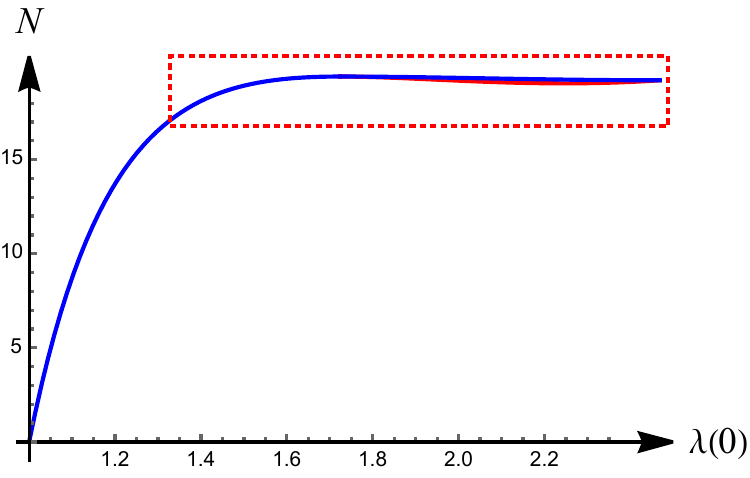}}
	\subfigure{\includegraphics[scale=0.6]{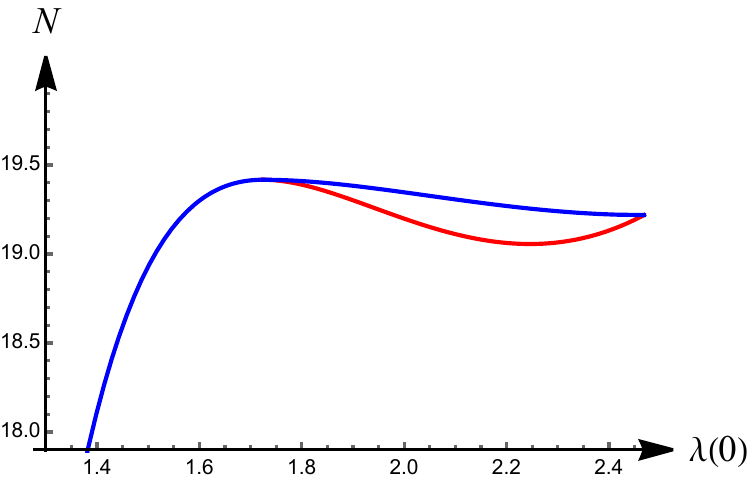}}
	\caption{The axial force $N$ versus the axial stretch $\lambda(0)$ as $J_\textrm{m}=20$, $\kappa=0.5$, and $\nu=10$. The right image is a magnified view. The homogeneous deformation \eqref{force} is depicted by the blue curve up to $\lambda_{\mathrm{cr}}=1.67381$, the remaining part is obtained using \eqref{amplitude}. The red curve, given by \eqref{force}, is unstable and shown for reference only.}\label{force-lambda}
\end{figure}

The left image in  Figure \ref{force-lambda} shows the axial force $N$ against the stretch $\lambda(0)$ for $J_{\mathrm m}=20, \kappa=0.5,\nu=10$ before the Maxwell stage, the right image is a magnified view. The homogeneous deformation \eqref{force} is depicted by the blue curve up to $\lambda_{\mathrm{cr}}=1.67381$, the remaining part is obtained using \eqref{amplitude}. The red curve, given by \eqref{force}, is unstable and shown for reference only \cite{deBotton2013,liu2023}. 

\subsection{Fixed axial length}\label{Fixed-length}

For the second loading scenario we keep the pre-stretch constant and use  $\nu$ as the loading parameter. It is shown in  \cite{Liu2023MMS} that, depending on the amount of pre-stretch, localized necking or bulging occurs for a residual stress larger than some critical value.  

In Figure  \ref{bifurcation-curve} we define $\lambda_{\mathrm m}=2.0567$ and show that an increase in residual stress with constant pre-stretch $\lambda<\lambda_{\mathrm m}$ initiates localized necking, with $\lambda>\lambda_{\mathrm m}$ localized bulging. We now use the one-dimensional model to validate this finding. 

The fixed length condition requires that by
\begin{equation}
\int_0^L\lambda(Z)\textrm{d}Z=\lambda_\textrm{f}L,
\label{fixed-length}
\end{equation}
where  the constant pre-stretch $\lambda_\textrm{f}>1$.  Applying the discretization procedure results in
\begin{equation}
\frac{1}{2}\lambda(0)+\sum_{j=1}^{k-1}\lambda(j)+\frac{1}{2}\lambda(k)-\frac{\lambda_\textrm{f}}{h}L=0,
\label{fixed-length-di}
\end{equation}
where $k$ is the number of nodes and  $h=L/k$ is the step size. To formulate a well-defined system,  two boundary conditions are required. Similar to the previous loading scenario, we neglect the exponentially small error \cite{Wang2020} and use
\begin{equation}
\lambda'(0)=0, \quad \lambda'(L)=0,
\label{right}
\end{equation}
which have the discretized forms
\begin{eqnarray}
\frac{\lambda_1-\lambda_{-1}}{2h}=0,\quad \frac{\lambda_{k+1}-\lambda_{k-1}}{2h}=0.
\label{bc-fixed-length}
\end{eqnarray}

As $\lambda(0)$ and $\lambda_\infty$ are both unknown, we rewrite  \eqref{Model-discretized} in the form 
\begin{equation}
G'(\lambda_j)-\frac{1}{2}D'(\lambda_j)\left(\frac{\lambda_{j+1}-\lambda_{j-1}}{2h}\right)^2-D(\lambda_j)\left(\frac{\lambda_{j+1}-2\lambda_j+\lambda_{j-1}}{h^2}\right)=0,\quad j=0,1,\cdots,k,
\label{Model-discretized-new}
\end{equation}
which, when combined with \eqref{fixed-length-di}, result in $(k+2)$ algebraic equations  for $\lambda_j, j=0,\dots,k$ and $\lambda_\infty$.

Consider a cylindrical solid with material extensibility limit $J_\textrm{m}=100$, $\kappa=0.5$ and constant pre-stretch $\lambda_\textrm{f}=1.9$. Bifurcation occurs when the residual stress parameter $\nu_\textrm{cr}=8.17618$ or, equivalently  $(\kappa\nu^2)_{\mathrm {cr}}=33.425$, see Figure \ref{bifurcation-curve}. In addition, it  shows that the axial pre-stretch $\lambda_\textrm{f}$ is less than the transition value $\lambda_\textrm{m}=2.0567$  indicating  localized necking. 

\begin{figure}[!ht]
\centering\includegraphics[scale=0.8]{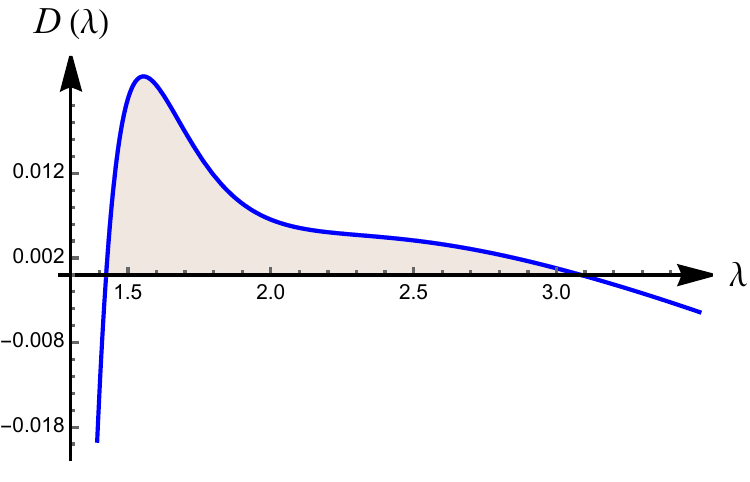}\caption{The dependence of $D(\lambda)$ on $\lambda$ when the material extensibility  $J_\textrm{m}=100$, $\kappa=0.5$ and $\nu=8.17618$. The two roots are located at $\lambda_\textrm{t1}=1.42367$ and $\lambda_\textrm{t2}=3.08483$.}
\label{Dlambda-fixedlength-necking}
\end{figure}

To use the one-dimensional model \eqref{1d-model} we first compute the changes of $D(\lambda)$  as a function of $\lambda$ and show the dependence in Figure \ref{Dlambda-fixedlength-necking}. The two roots are calculated as  $\lambda_\textrm{t1}=1.42367$ and $\lambda_\textrm{t2}=3.08483$  with $\lambda_\textrm{t1}<\lambda_\textrm{f}<\lambda_\textrm{t2}$ such that model \eqref{1d-model}  admits a localized solution. 

\begin{figure}[!ht]
	\centering\includegraphics[scale=0.6]{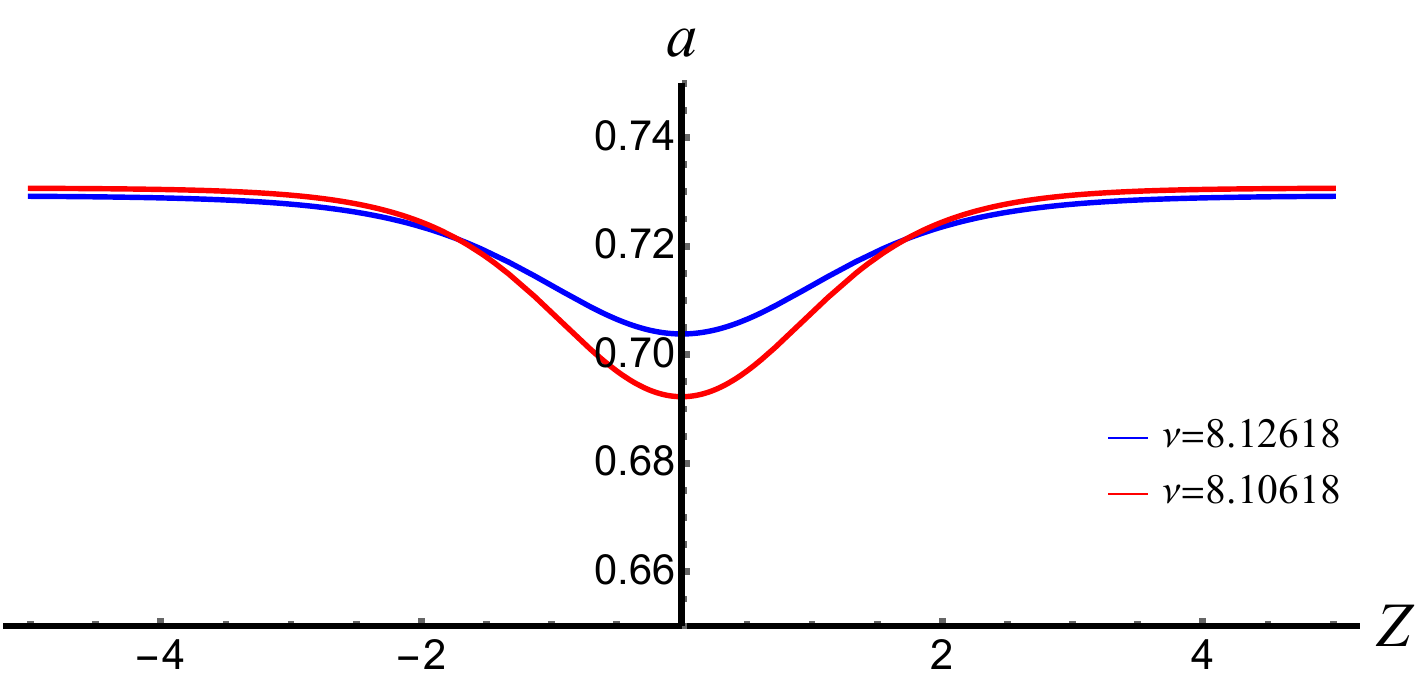}\caption{The blue and red curves represent the post-bifurcation profiles of a solid cylinder with constant  pre-stretch $\lambda_\textrm{f}=1.9$,  $J_\textrm{m}=100$, $\kappa=0.5$. The residual stress parameters are $\nu=8.12618$ and $\nu=8.10618$, respectively. }\label{necking-profile-Jm100}
\end{figure}

For a subcritical bifurcation, the loading parameter after bifurcation decreases and we take $\nu=\nu_\textrm{cr}-0.05=8.12618$. To find a solution, we use the starting value $\lambda_\infty=1.899$, which is slightly lower than $\lambda_\textrm{f}$ and obtain $\lambda_0$ from \eqref{amplitude}.  Then, we numerically solve \eqref{1d-model} with \eqref{fixed-length} and  adjust the value of $\lambda_\infty$ until equation \eqref{fixed-length} is satisfied. The profile of the solid cylinder  in the post-bifurcation state is  given by $a(Z)=\lambda(Z)^{-1/2}$ and is depicted in Figure \ref{necking-profile-Jm100}. The solution clearly shows that necking has occurred. 

\begin{figure}[!ht]
	\centering\includegraphics[scale=0.8]{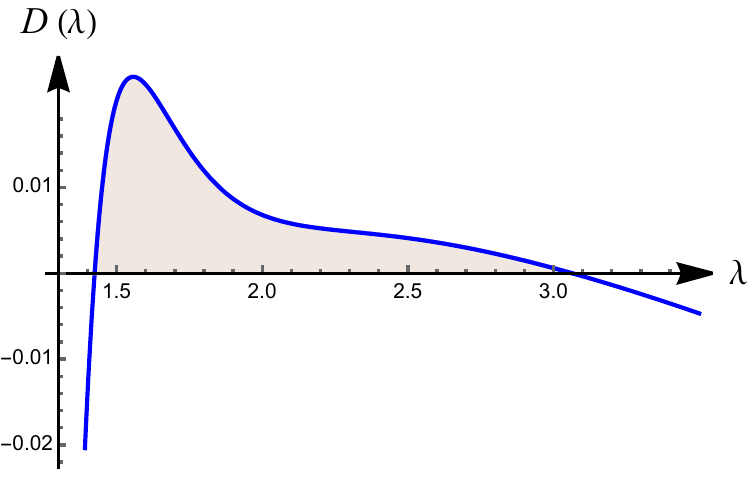}\caption{The dependence of $D(\lambda)$ on $\lambda$ as $J_\textrm{m}=100$, $\kappa=0.5$, and $\nu=8.09505$.  The two roots are located at $\lambda_\textrm{t1}=1.42471$ and $\lambda_\textrm{t2}=3.06389$.}
	\label{Dlambda-fixedlength-bulging}
\end{figure}

For the last illustration, we consider a cylindrical solid with constant pre-stretch $\lambda_{\mathrm f}> \lambda_{\mathrm m}$, where $\lambda_{\mathrm m}=2.0567$. Figure  \ref{bifurcation-curve} shows that  an increase in residual stress  initiates localized bulging. For  $J_\textrm{m}=100, \kappa=0.5$  and $\lambda_\textrm{f}=2.2$  localized bifurcation occurs when  $\nu_\textrm{cr}=8.09505$. Figure \ref{Dlambda-fixedlength-bulging} shows the dependence of  $D(\lambda)$ on $\lambda$ and the two roots located at $\lambda_\textrm{t1}=1.42471$ and $\lambda_\textrm{t2}=3.06389$. Hence, the one-dimensional model \eqref{1d-model} admits a localized solution for a constant pre-stretch $\lambda_\textrm{t1}<\lambda_{\mathrm f} <\lambda_\textrm{t2}$.

\begin{figure}[!ht]
	\centering\includegraphics[scale=0.6]{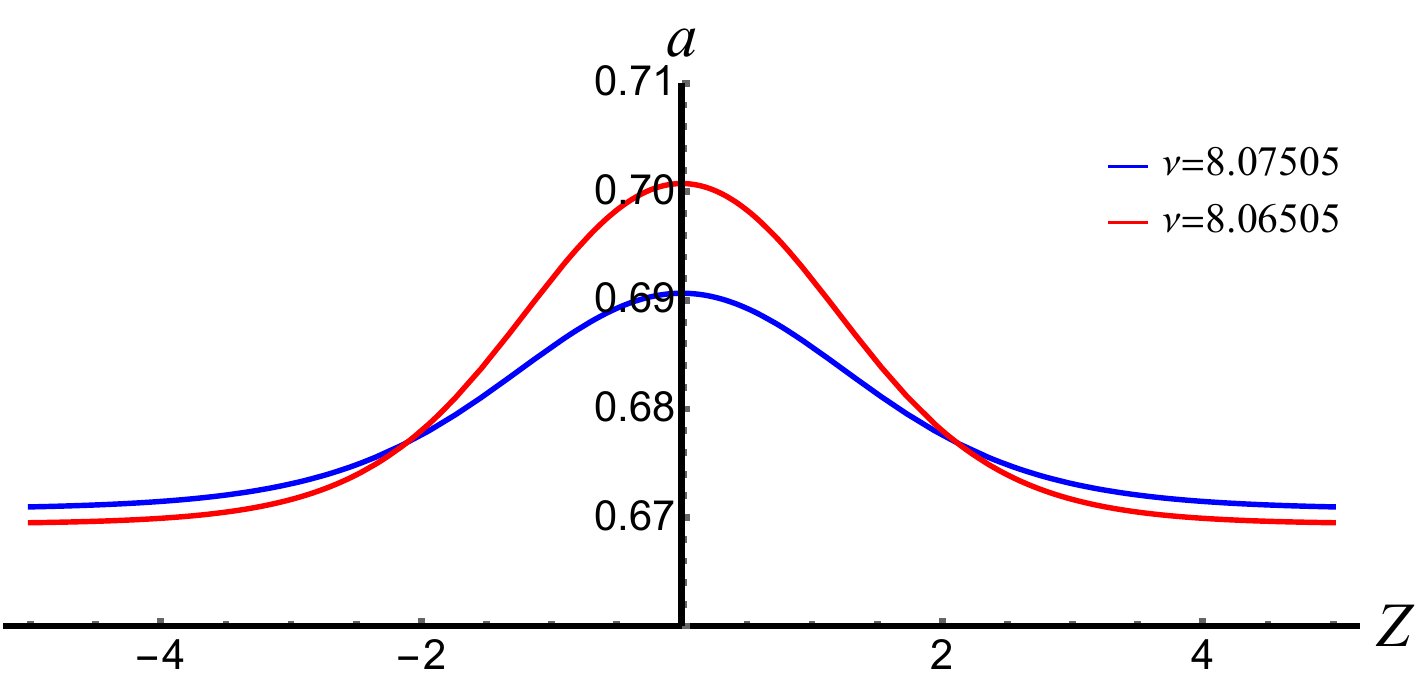}\caption{The blue and red curves represent the bulging profiles of a solid cylinder with constant  pre-stretch $\lambda_\textrm{f}=2.2$,  $J_\textrm{m}=100$, $\kappa=0.5$. The residual stress parameters are $\nu=8.07505$ and $\nu=8.06505$, respectively. }\label{bulging-profile-Jm100}
\end{figure}

Figure \ref{bulging-profile-Jm100} displays the profiles of the cylinder for different values of the residual stress parameter $\nu$. It clearly shows that bulging has occurred, validating the results obtained in the linear bifurcation analysis \cite{Liu2023MMS}.

\clearpage

\section{Conclusions}\label{Conclusions}

Instabilities in the form of periodic or localized  pattern   in residually stressed materials have been observed in biological structures and in man-made devices. In particular, the residual stress in biological materials  arises due to growth and remodeling and has a signifiant influence on the mechanical response, as shown in, for example,  \cite{Holzapfel2000}. In manufacturing, residual stress may improve material performance or cause weaknesses and possibly a reduction in the life of components \cite{Melnikov2021}. 

The bifurcation behavior leading to the formation of periodic pattern  is based on the theory of incremental deformations superimposed on a known finitely deformed configuration  \cite{Haughton1979a,Haughton1979b}.  Contrasting, the analysis  of localized instabilities, such as necking or bulging, focuses on initiation,  growth and propagation. They  play a role in nanotubes that bridge neighboring cells \cite{Veranic2008} and in trauma induced  beading instabilities responsible for damage of axons \cite{Hemphill2015}. Recently, the effect of surface tension to initiate localized instabilities in soft solids has been recognized \cite{Fu2021}, and the one-dimensional theory can be used to better understand this phenomenon. The theory presented in this paper can also be extended to evaluate the effect of residual stress on localized pattern formation in liquid crystal elastomers \cite{li2022lce} and in the formation and propagation of  abdominal aortic aneurysms \cite{Ahamed2016}.

We  focused on a residually stressed Gent material  and investigated localized bifurcation of a circular cylindrical solid subject to axial stretch. We discussed the dimensional reduction of the total potential energy functional and used the resulting one-dimensional model to analyze the post-bifurcation behavior.

Two loading scenarios were considered. In the first, we keep the residual stress constant and monotonically  increase the axial stretch until localized bifurcation occurs. In the second, we apply a constant pre-stretch  and use the residual stress as the loading parameter. 

The bifurcation diagram of the first loading sequence shows that the deformation is  homogeneous until the critical stretch  is reached. This is followed by the non-trivial solution during which the axial stretch of the unstable region increases compared to the far end value. We plot the radial deformation as a function of the axial position and show that localized necking occurs. We also optimize the material extensibility to show that the solution of the one-dimensional model is valid until the Maxwell values of stretch are reached.

For the  second loading sequence we determine the transition stretch $\lambda_{\mathrm m}$, which distinguishes  between necking or bulging modes. Specifically, we find that an increase in residual stress with constant stretch $\lambda<\lambda_{\mathrm m}$ initiates localized necking, an axial stretch  $\lambda>\lambda_{\mathrm m}$ leads to localized bulging. The corresponding axial profiles of the axisymmetric solid  in the post-bifurcation state are shown.  

\section*{Acknowledgment}
We thank Prof. Yibin Fu from Keele University for  helpful discussions and valuable suggestions. 

\section*{Funding}

\noindent This work was supported by the National Natural Science Foundation of China (Project Nos 12072227 and 12021002). 

\section*{ORCID iD}

\noindent 
Yang Liu https://orcid.org/0000-0001-9517-3833\\
Xiang Yu https://orcid.org/0000-0002-3378-6340\\
Luis Dorfmann    https://orcid.org/0000-0002-9665-0272


\end{document}